\pdfoutput=1
\documentclass[twocolumn]{jpsj3}
\usepackage{bm}
\usepackage{tabularx}
\usepackage{ascmac}
\usepackage{amsmath}
\usepackage{wrapfig}
\usepackage{graphicx}
\usepackage{float}
\usepackage[FIGBOTCAP]{subfigure}
\usepackage{color}

\bmdefine{\bdi}{i}
\bmdefine{\bdj}{j}
\bmdefine{\bdx}{x}
\bmdefine{\bdy}{y}
\bmdefine{\bdr}{r}
\bmdefine{\bdR}{R}
\bmdefine{\bdS}{S}
\bmdefine{\bdL}{L}
\bmdefine{\bds}{s}
\bmdefine{\bdl}{l}
\bmdefine{\bdJ}{J}
\bmdefine{\bdA}{A}
\bmdefine{\bdE}{E}
\bmdefine{\bdD}{D}
\bmdefine{\bdQ}{Q}
\bmdefine{\bdq}{q}
\bmdefine{\bdk}{k}
\bmdefine{\bdzero}{0}
\bmdefine{\bdv}{v}
\bmdefine{\bde}{e}
\bmdefine{\bdb}{b}
\bmdefine{\bddelta}{\delta}

\title{Light-Induced Mirror Symmetry Breaking and Charge Transport}

\author{Naoya Arakawa$^{1}$\thanks{E-mail address: arakawa@phys.chuo-u.ac.jp},
Kenji Yonemitsu$^{1,2}$}
\inst{$^{1}$The Institute of Science and Engineering,
  Chuo University, Bunkyo, Tokyo, 112-8551, Japan\\
  $^{2}$Department of Physics,
  Chuo University, Bunkyo, Tokyo 112-8551, Japan}

\abst{
  We propose that light can break mirror
  symmetries and combining symmetries with a uniform time translation,
  and their breaking is characterized by an off-diagonal charge conductivity.
  Taking periodically driven graphene as an example, 
  we show that
  mirror symmetries about the $xz$ and $yz$ planes
  and the combining symmetries,
  the symmetries of combinations of the mirror operations about these planes
  and a uniform time translation,
  can be broken
  by linearly or circularly polarized light.
  We also show that 
  this symmetry breaking induces the time-averaged
  off-diagonal symmetric or antisymmetric
  charge conductivity in a nonequilibrium steady state
  with linearly or circularly, respectively, polarized light.
  Our results are experimentally testable in pump-probe measurements.
  This work will pave the way for controlling mirror symmetries via light and
  utilizing the light-induced mirror symmetry breaking.   
}

\begin{document}
\maketitle

\section{Introduction}

Light can break symmetries in time and space. 
For example, circularly polarized light (CPL) can break
the time-reversal symmetry~\cite{Trev1,Trev2,Trev3}.
If CPL is applied to a non-magnetic material,
it can induce the magnetization~\cite{IFE1};
the direction of this light-induced magnetization
can be reversed by changing the helicity of CPL~\cite{IFE2}.
CPL can also induce
the anomalous Hall effect (AHE)~\cite{Oka-PRB,Light-AHE-exp2},
in which 
a charge current perpendicular to an applied electric field 
is generated~\cite{AHE-Hall,Karplus-Luttinger,AHE-review};
the magnitude and direction of this current can be changed by
varying the amplitude and helicity of CPL~\cite{Oka-PRB,Light-AHE-exp2,Mikami,NA-FloquetSHE}.
Then,
bicircularly polarized light~\cite{Bicirc-NatPhoto,Bicirc-Oka},
which consists of a linear combination of
left-handed and right-handed CPL, can break
not only the time-reversal,
but also the inversion symmetry~\cite{Bicirc-PRL}.
In fact, it can be used to realize
noncentrosymmetric magnetic topological phases~\cite{Bicirc-PRL}
and generate electric polarization~\cite{Bicirc-PTEP}. 
Since the application of light enables us
to engineer electronic, magnetic, or transport properties
without changing materials,
it is crucial to understand
which symmetry is broken by light
and how its symmetry breaking affects the properties. 

In this paper, 
we show the mirror symmetry breaking by CPL or by linearly polarized light (LPL),
which results in an off-diagonal
antisymmetric or symmetric,
respectively,
charge conductivity
(i.e., $\sigma_{xy}^{\textrm{C}}=-\sigma_{yx}^{\textrm{C}}$
or $\sigma_{xy}^{\textrm{C}}=\sigma_{yx}^{\textrm{C}}$, respectively).
This is demonstrated for periodically driven graphene.
The difference between the cases with CPL and LPL 
comes from the difference in the time-reversal symmetry.
The main results are summarized in Table \ref{tab}.
Our results suggest
that
a combination of time-reversal symmetry breaking
and mirror symmetry breaking
is the origin of the light-induced AHE,
and that
the off-diagonal symmetric charge conductivity could be used to
detect whether mirror symmetries are broken or preserved
in the presence of the time-reversal symmetry.

\begin{table}
  \caption{\label{tab}
    Properties of systems driven by CPL or LPL.
    The difference among LPL1, LPL2, and LPL3 is
    about the polarization:
    $A_{x}\neq 0$ and $A_{y}\neq 0$ in LPL1;
    $A_{x}\neq 0$ and $A_{y}=0$ in LPL2;
    and $A_{x}=0$ and $A_{y}\neq 0$ in LPL3.
    $T_{\textrm{rev}}$ represents the time-reversal symmetry,
    $\sigma_{\textrm{m}}$ or $\sigma_{\textrm{m}}^{\prime}$
    represents the mirror symmetry about the $xz$ or $yz$ plane, respectively,
    and $C_{3}$ represents the $C_{3}$ rotational symmetry around the $z$ axis.
    $\sigma_{\textrm{m}}T_{t}$, $\sigma_{\textrm{m}}^{\prime}T_{t}$,
    or $C_{3}T_{t}$ represents 
    the symmetry of a combination of
    the mirror operation about the $xz$ or $yz$ plane
    or the $C_{3}$ rotation operation 
    and a uniform time translation $T_{t}$.
    $\sigma_{yx}^{\textrm{C}}$ represents
    an off-diagonal charge conductivity.
    $\sigma_{\textrm{m}}T_{t}$ or $\sigma_{\textrm{m}}^{\prime}T_{t}$
    is preserved with LPL3 or LPL2, respectively,
    if $T_{t}: t \rightarrow t-\frac{\pi}{\Omega}$. 
    $C_{3}T_{t}$ is preserved with CPL
    if $T_{t}: t \rightarrow t+\frac{2\pi}{3\Omega}$.
    \vspace{8pt}
  }
  \begin{center}
  \begin{tabular}{ccccc}
    \hline
      & CPL & LPL1 & LPL2 & LPL3 \\[2pt] \hline\\[-6pt]    
      $T_{\textrm{rev}}$ & Broken & Preserved & Preserved & Preserved\\[4pt]
      $\sigma_{\textrm{m}}$ & Broken & Broken & Preserved & Broken\\[4pt]
      $\sigma_{\textrm{m}}T_{t}$ & Broken & Broken
      & Preserved & Preserved\\[4pt]
      $\sigma_{\textrm{m}}^{\prime}$ & Broken & Broken & Broken & Preserved\\[4pt]
      $\sigma_{\textrm{m}}^{\prime}T_{t}$ & Broken & Broken
      & Preserved & Preserved\\[4pt]
      $C_{3}$ & Broken & Broken
      & Broken & Broken\\[4pt]
      $C_{3}T_{t}$ & Preserved & Broken
      & Broken & Broken\\[4pt]
    $\sigma_{yx}^{\textrm{C}}$ & Antisymmetric  & Symmetric & Vanishing & Vanishing\\[3pt]
    \hline
  \end{tabular}
  \end{center}
\end{table}

\section{Model}

Our periodically driven electron system is described by the Hamiltonian,
\begin{align}
  H=H_{\textrm{s}}(t)+H_{\textrm{b}}+H_{\textrm{sb}}.\label{eq:H}
\end{align}
Here $H_{\textrm{s}}(t)$ is
the Hamiltonian of the system driven by a light field $\bdA(t)$,
the effect of which is treated as the Peierls phase factors:
\begin{align}
  H_{\textrm{s}}(t)
  &=\sum_{\bdk}\sum_{a,b=A,B}\sum_{\sigma=\uparrow,\downarrow}
  \epsilon_{ab}(\bdk,t)
  c_{\bdk a\sigma}^{\dagger}c_{\bdk b\sigma},\label{eq:Hs}
\end{align}
where
$\epsilon_{AB}(\bdk,t)=\epsilon_{BA}(\bdk,t)^{\ast}
=t_{\textrm{NN}}\sum_{l=0}^{2}e^{-i[\bdk+e\bdA(t)]\cdot\bdR_{l}}$,
$\epsilon_{aa}(\bdk,t)=0$,
$\bdR_{0}={}^{t}(0\ 1)$,
$\bdR_{1}={}^{t}(-\frac{\sqrt{3}}{2}\ -\frac{1}{2})$,
$\bdR_{2}={}^{t}(\frac{\sqrt{3}}{2}\ -\frac{1}{2})$,
$t_{\textrm{NN}}$ is the hopping integral
between nearest neighbor sites on a honeycomb lattice~\cite{Graphene-review} 
without $\bdA(t)$, 
and $c_{\bdk a\sigma}^{\dagger}$ and $c_{\bdk a\sigma}$
are the creation and annihilation operators of an electron
for momentum $\bdk$, sublattice $a$, and spin $\sigma$.
Hereafter, 
we set $\hbar=c=k_{\textrm{B}}=a_{\textrm{NN}}=1$,
where $a_{\textrm{NN}}$ is the length between nearest neighbor sites.
Then,
$H_{\textrm{b}}$ is the Hamiltonian of
the Buttiker-type heat bath~\cite{HeatBath1,HeatBath2},
which is in equilibrium at temperature $T$:
$H_{\textrm{b}}=\sum_{i}\sum_{p}(\epsilon_{p}-\mu_{\textrm{b}})b_{ip}^{\dagger}b_{ip}$,
where $b_{ip}$ and $b_{ip}^{\dagger}$ are the annihilation and creation operators
of a bath's fermion at site $i$ for mode $p$,
and $\epsilon_{p}$ and $\mu_{\textrm{b}}$
are the energy and chemical potential of a bath's fermion;
$\mu_{\textrm{b}}$ is determined from the condition that there is no current between
the system and bath.
In addition,
$H_{\textrm{sb}}$ is the system-bath coupling Hamiltonian~\cite{Tsuji,Mikami,NA-FloquetSHE}:
$H_{\textrm{sb}}=\sum_{i}\sum_{p}\sum_{a=A,B}\sum_{\sigma=\uparrow,\downarrow}V_{pa\sigma}(c_{ia\sigma}^{\dagger}b_{ip}+b_{ip}^{\dagger}c_{ia\sigma})$,
where $V_{pa\sigma}$ is the system-bath coupling constant.

We have considered $H_{\textrm{b}}$ and $H_{\textrm{sb}}$, as well as $H_{\textrm{s}}(t)$,
because the damping due to the system-bath coupling makes
the system a nonequilibrium steady state~\cite{Tsuji,Mikami,NA-FloquetSHE}.
Such a relaxation mechanism is necessary for periodically driven systems,
in which the heating due to the driving field exists~\cite{Heating1,Heating2}. 

\section{Light-induced mirror symmetry breaking}

\begin{figure}
  \includegraphics[width=86mm]{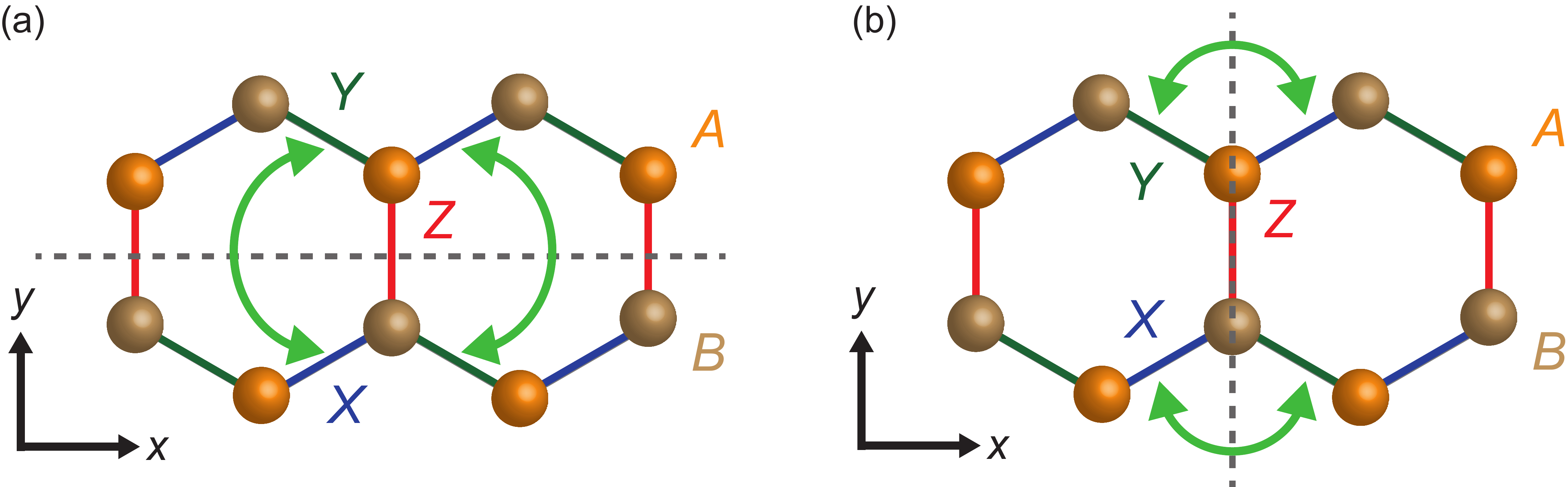}
  \caption{\label{fig1}
    (Color online) The honeycomb lattice and (a) the $xz$ or (b) the $yz$ mirror plane.
    The dashed lines denote the mirror planes.
    The green arrows represent the bonds which are connected
    by the mirror symmetry.
    $A$ or $B$ represents sublattice $A$ or $B$, respectively.
    The red, blue, and
    dark green bonds represent
    $Z$, $X$, and $Y$ bonds, respectively. 
    The $x$ and $y$ axes are also drawn.
  }
\end{figure}

First, we analyze
the polarization dependence of the light-induced mirror symmetry breaking.
For our periodically driven electron system, 
whether a mirror symmetry
is preserved or broken is determined
by the symmetry of the kinetic energy, which
is characterized by
\begin{align}
  \epsilon_{AB}(\bdk,t)
  &= t_{AB}^{Z}(t)e^{-ik_{y}}
  +t_{AB}^{X}(t)e^{i\frac{\sqrt{3}}{2}k_{x}}e^{i\frac{k_{y}}{2}}\notag\\
  &+t_{AB}^{Y}(t)e^{-i\frac{\sqrt{3}}{2}k_{x}}e^{i\frac{k_{y}}{2}},\label{eq:e_AB-App}\\
  \epsilon_{BA}(\bdk,t)
  &= t_{BA}^{Z}(t)e^{ik_{y}}
  +t_{BA}^{X}(t)e^{-i\frac{\sqrt{3}}{2}k_{x}}e^{-i\frac{k_{y}}{2}}\notag\\
  &+t_{BA}^{Y}(t)e^{i\frac{\sqrt{3}}{2}k_{x}}e^{-i\frac{k_{y}}{2}},\label{eq:e_BA-App}
\end{align}
where $t_{AB}^{Z}(t)=t_{BA}^{Z}(t)^{\ast}$,
$t_{AB}^{X}(t)=t_{BA}^{X}(t)^{\ast}$,
and $t_{AB}^{Y}(t)=t_{BA}^{Y}(t)^{\ast}$.
Here
$Z$, $X$, and $Y$ represent the three bonds between nearest neighbor sites
(see Fig. \ref{fig1}).
If the hopping integrals satisfy
\begin{align}
  t_{AB}^{Z}(t)=t_{BA}^{Z}(t),
  t_{AB}^{Y}(t)=t_{BA}^{X}(t),
  t_{AB}^{X}(t)=t_{BA}^{Y}(t),\label{eq:sym-xzMirror}
\end{align}
the mirror symmetry about the $xz$ plane [Fig. \ref{fig1}(a)] is preserved;
otherwise, it is broken.

We begin with the system driven by LPL. 
The field of LPL is described by 
\begin{align}
  \bdA_{\textrm{pump}}(t)={}^{t}(A_{0}\alpha_{x}\cos\Omega t\ A_{0}\alpha_{y}\cos\Omega t),
  \label{eq:A_pump-LPL}
\end{align}
where $\Omega=2\pi/T_{\textrm{p}}$ is the light frequency,
and $T_{\textrm{p}}$ is the period of $\bdA_{\textrm{pump}}(t)$.
In this case, 
$t_{AB}^{Z}(t)
=t_{\textrm{NN}}e^{-iu\alpha_{y}\cos\Omega t}$,
$t_{AB}^{X}(t)
=t_{\textrm{NN}}e^{iu\alpha_{x}\frac{\sqrt{3}}{2}\cos\Omega t}e^{iu\alpha_{y}\frac{1}{2}\cos\Omega t}$,
and
$t_{AB}^{Y}(t)
=t_{\textrm{NN}}e^{-iu\alpha_{x}\frac{\sqrt{3}}{2}\cos\Omega t}e^{iu\alpha_{y}\frac{1}{2}\cos\Omega t}$,
where $u=eA_{0}$.
For the LPL with $\alpha_{x}\neq 0$, $\alpha_{y}=0$,
these hopping integrals satisfy Eq. (\ref{eq:sym-xzMirror}),
which means that
the mirror symmetry about the $xz$ plane is preserved.
Meanwhile,
for the LPL with $\alpha_{x}=0$, $\alpha_{y}\neq 0$ or
with $\alpha_{x}\neq 0$, $\alpha_{y}\neq 0$,
this mirror symmetry is broken.

We turn to the case with CPL.
The field of CPL is given by
\begin{align}
  \bdA_{\textrm{pump}}(t)
  ={}^{t}(A_{0}\cos\Omega t\ A_{0}\sin\Omega t).
  \label{eq:A_pump-CPL}
\end{align}
In a similar way,
we can show that
the mirror symmetry about the $xz$ plane is broken.

\begin{figure}
  \includegraphics[width=86mm]{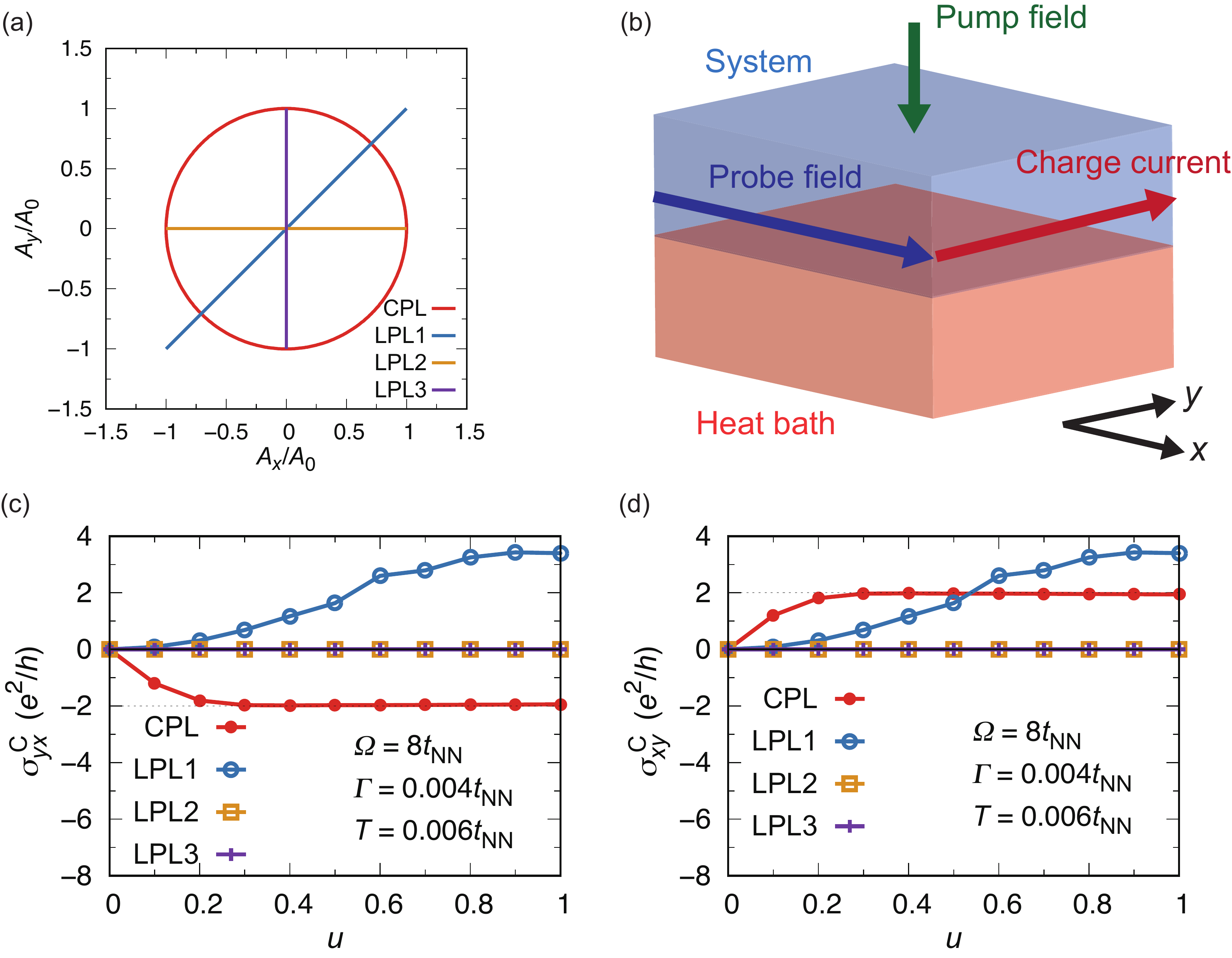}
  \caption{\label{fig2}
    (Color online)
    (a) The trajectories of $\bdA_{\textrm{pump}}(t)$
    for CPL, LPL1, LPL2, and LPL3.
    (b) The setup for the pump-probe measurements of $\sigma_{yx}^{\textrm{C}}$ in
    our periodically driven system.
    For the setup of $\sigma_{xy}^{\textrm{C}}$,
    the directions of the probe field and charge current are interchanged.
    The arrow of the pump light indicates the direction of travel,
      and that of the probe light indicates the direction of
      the component of the corresponding electric field.
    (c), (d) $\sigma_{yx}^{\textrm{C}}$ and $\sigma_{xy}^{\textrm{C}}$ as functions of $u=eA_{0}$
    for graphene driven by CPL, LPL1, LPL2, and LPL3.
    The horizontal dashed lines in (c) and (d)
    correspond to $-2e^{2}h^{-1}$ and $2e^{2}h^{-1}$, respectively. 
  }
\end{figure}

We make five remarks.
First, 
we can similarly show that
the mirror symmetry about the $yz$ plane [Fig. \ref{fig1}(b)] is broken
by LPL with $\alpha_{x}\neq 0$, $\alpha_{y}=0$,
by that with $\alpha_{x}\neq 0$, $\alpha_{y}\neq 0$,
and by CPL,
whereas it is preserved by LPL with $\alpha_{x}=0$, $\alpha_{y}\neq 0$.
Second,
the same polarization dependence holds for the Floquet Hamiltonian 
(see Appendix A).
Third,
the similar arguments can be used to discuss
whether a mirror symmetry is broken or not
in the other periodically driven electron systems.
Fourth,
a mirror symmetry of a periodically driven electron system 
does not necessarily match that of the trajectory of $\bdA_{\textrm{pump}}(t)$.
For example,
CPL breaks the mirror symmetry about the $xz$ plane,
whereas the trajectory of its $\bdA_{\textrm{pump}}(t)$
has the mirror symmetry in the $A_{x}$-$A_{y}$ plane
[see Fig. \ref{fig2}(a)].
Fifth,
our mirror symmetry,
the symmetry about a mirror operation in crystals,
is essentially different from
a symmetry discussed in Ref.~\cite{Fiete},
the symmetry about the energy spectrum as a function of magnetic flux.
Namely, the light-induced symmetry breaking discussed in Ref.~\cite{Fiete}
is not about crystal's mirror symmetry.
To the best of our knowledge,
our paper is the first work demonstrating
the light-induced breaking of crystal's mirror symmetry.

The above arguments show that
the mirror symmetries about the $xz$ and $yz$ planes are broken 
with CPL or LPL1 (i.e., LPL with finite $\alpha_{x}$ and $\alpha_{y}$).
To study a time-averaged quantity in a nonequilibrium steady state,
we need to discuss 
not only the mirror symmetries,
but also its combining symmetries,
the symmetries of combinations of the mirror operations about the $xz$ and $yz$ planes
and a uniform time translation,
because that quantity is not affected by such a translation.
(Note that such a combining symmetry is sometimes called
a space-time or dynamical symmetry.)
In general, there is a case that 
a spatial symmetry is broken, but its combining symmetry is preserved;
in such a case, a time-averaged quantity in a nonequilibrium steady state behaves
as if the spatial symmetry were preserved.  
As we show in Appendix B,
the combining symmetries for the mirror operations are also broken with CPL or LPL1. 
This contrasts with the combining symmetry for the $C_{3}$ rotation
in graphene driven by CPL (see Apppendix B):
the $C_{3}$ rotational symmetry is broken, but 
the symmetry of a combination of the $C_{3}$ rotation and
the uniform time translation $T_{t}: t \rightarrow t+\frac{2\pi}{3\Omega}$
is preserved~\cite{Mikami}.
This combining symmetry may be called a time-screw symmetry. 
In appendix B,
we also show that 
the combining symmetry for the mirror operation about
the $yz$ or $xz$ plane
is preserved with 
LPL2 (i.e., LPL with $\alpha_{x}\neq 0$, $\alpha_{y}=0$) 
or LPL3 (i.e., LPL with $\alpha_{x}=0$, $\alpha_{y}\neq 0$),
respectively, if $T_{t}: t \rightarrow t-\frac{\pi}{\Omega}$.
This combining symmetry may be called a time-glide symmetry.
We do not necessarily call the combining symmetry for a mirror operation a time-glide one 
because an analogy with an axial glide symmetry suggests that
a time-glide symmetry consists of a mirror operation and
the uniform time translation with $T_{\textrm{p}}/2$. 
Then, 
the $C_{3}$ rotational symmetry and its combining symmetry are both broken with LPL
(see Appendix B).
These results are summarized in Table \ref{tab}.

\section{Charge transport induced by mirror symmetry breaking}

Next, we study
the effects of the light-induced mirror symmetry breaking on transport properties.
To do this,
we use the Floquet linear-response theory~\cite{NA-FloquetSHE,Mikami,Eckstein}
for pump-probe measurements [Fig. \ref{fig2}(b)].
In this theory,
we set $\bdA(t)=\bdA_{\textrm{pump}}(t)+\bdA_{\textrm{prob}}(t)$
and treat the effects of $\bdA_{\textrm{pump}}(t)$ in the Floquet theory~\cite{Floquet1,Floquet2} 
and those of $\bdA_{\textrm{prob}}(t)$
in the linear-response theory~\cite{Kubo}.  
The $\bdA_{\textrm{pump}}(t)$ for LPL or CPL is given
by Eq. (\ref{eq:A_pump-LPL}) or (\ref{eq:A_pump-CPL}), respectively.  
Note that $\bdA_{\textrm{pump}}(t)$ is used to periodically drive the system,
whereas $\bdA_{\textrm{prob}}(t)$ is used to analyze its properties~\cite{Opt-review}.
In this theory,
we use the Floquet Hamiltonian for $H_{\textrm{s}}(t)$,
which is distinct from the Hamiltonian obtained
in a high-frequency expansion.
Using the Floquet linear-response theory,
we obtain a time-averaged charge conductivity
$\sigma_{\mu\nu}^{\textrm{C}}$
in the nonequilibrium steady state~\cite{NA-FloquetSHE,Mikami}
(see Appendix C),
\begin{align}
  &\sigma_{\mu\nu}^{\textrm{C}}
  =\frac{e^{2}}{V}\sum_{\bdk}\sum_{a,b,c,d=A,B}\sum_{\sigma,\sigma^{\prime}=\uparrow,\downarrow}
  \int_{-\Omega/2}^{\Omega/2}\frac{d\omega^{\prime}}{2\pi}
  \notag\\
  &\times
  \textrm{tr}
  \Bigl[v_{ab}^{\mu}(\bdk)
        \frac{\partial G_{b\sigma c\sigma^{\prime}}^{\textrm{R}}(\bdk,\omega^{\prime})}
             {\partial \omega^{\prime}}
        v_{cd}^{\nu}(\bdk)
        G_{d\sigma^{\prime}a\sigma}^{<}(\bdk,\omega^{\prime})\notag\\
  &\ \ \ -v_{ab}^{\mu}(\bdk)
          G_{b\sigma c\sigma^{\prime}}^{<}(\bdk,\omega^{\prime})
          v_{cd}^{\nu}(\bdk)
          \frac{\partial G_{d\sigma^{\prime}a\sigma}^{\textrm{A}}(\bdk,\omega^{\prime})}
               {\partial \omega^{\prime}}\Bigr],\label{eq:sig^C}
\end{align}
where
the trace is taken over the Floquet states
[i.e., $\textrm{tr}(ABCD)=\sum_{m,l,n,q=-\infty}^{\infty}A_{ml}B_{ln}C_{nq}D_{qm}$
with Floquet indices $m$, $l$, $n$, and $q$],
$V=\frac{N}{2}\frac{3\sqrt{3}}{2}$, 
$N$ is the number of sites, 
$[v_{ab}^{\nu}(\bdk)]_{mn}$ is the group velocity in the Floquet representation,  
and $[G_{a\sigma b\sigma^{\prime}}^{\textrm{R}}(\bdk,\omega^{\prime})]_{mn}$,
$[G_{a\sigma b\sigma^{\prime}}^{\textrm{A}}(\bdk,\omega^{\prime})]_{mn}$,
and $[G_{a\sigma b\sigma^{\prime}}^{<}(\bdk,\omega^{\prime})]_{mn}$
are the retarded, advanced, and lesser Green's functions, respectively,
in the Floquet representation.
(For more details, see Appendix C.) 
These Green's functions are determined from Dyson's equation
with the damping $\Gamma$ due to the second-order perturbation of $H_{\textrm{sb}}$
(see Appendix D).
Note that
$\sigma_{\mu\nu}^{\textrm{C}}$ is equivalent to the anomalous Hall conductivity
if and only if it is antisymmetric. 

Using Eq. (\ref{eq:sig^C}),
we numerically evaluate $\sigma_{yx}^{\textrm{C}}$ and $\sigma_{xy}^{\textrm{C}}$
for graphene driven by CPL, LPL1, LPL2, and LPL3.
(For details of the numerical calculations, see Appendix E.)
The directions of the probe field and the observed charge current are fixed: 
for $\sigma_{yx}^{\textrm{C}}$ (or $\sigma_{xy}^{\textrm{C}}$),
the charge current along the $y$ (or $x$) axis
is generated with the probe field applied along the $x$ (or $y$) axis.
CPL is described by Eq. (\ref{eq:A_pump-CPL}),
and LPL1, LPL2, or LPL3 is described by Eq. (\ref{eq:A_pump-LPL})
with $\alpha_{x}=\alpha_{y}=1$, with $\alpha_{x}=1$ and $\alpha_{y}=0$,
or with $\alpha_{x}=0$ and $\alpha_{y}=1$, respectively [Fig. \ref{fig2}(a)];
as described above,
the mirror symmetry about the $xz$ or $yz$ plane is preserved
only for LPL2 or LPL3, respectively.
We set
$\Omega=8t_{\textrm{NN}}$
and $t_{\textrm{NN}}=1$;
our light is off-resonant, i.e., $\Omega > W$,
where $W(=6t_{\textrm{NN}})$ is the bandwidth without light.
(We have summarized the main results in Table \ref{tab}.)
Except for the results of
the damping or temperature dependence of $\sigma_{yx}^{\textrm{C}}$ with LPL1,
we set $\Gamma=0.004t_{\textrm{NN}}$ and $T=0.006t_{\textrm{NN}}$.
When discussing the damping dependence,
we set $T=0.006t_{\textrm{NN}}$
and compare the results obtained at
$\Gamma=0.004t_{\textrm{NN}}$, $0.002t_{\textrm{NN}}$,  and $0.006t_{\textrm{NN}}$;
when discussing the temperature dependence, 
we set $\Gamma=0.004t_{\textrm{NN}}$
and compare the results obtained at
$T=0.006t_{\textrm{NN}}$, $0.004t_{\textrm{NN}}$, and $0.008t_{\textrm{NN}}$.

Figure \ref{fig2}(c) shows the $u$ dependences of $\sigma_{yx}^{\textrm{C}}$
in graphene driven by CPL, LPL1, LPL2, and LPL3.
$\sigma_{yx}^{\textrm{C}}$ for $u\neq 0$ is finite for CPL and LPL1,
whereas it vanishes for LPL2 and LPL3.
The similar results are obtained also
for $\sigma_{xy}^{\textrm{C}}$ [Fig. \ref{fig2}(d)].
These results are consistent
with the properties of the mirror symmetries about the $xz$ and $yz$ planes
and their combining symmetries (see Table \ref{tab}).
Therefore,
the mirror symmetries and their combining symmetries play a vital role in discussing
the off-diagonal charge conductivities.  
Note that since $u=eA_{0}=eE_{0}/\Omega$ is dimensionless,
the $u$ dependence of $\sigma_{yx}^{\textrm{C}}$ at fixed $\Omega$
gives its dependence on $E_{0}$, the amplitude of the light field.

One of the main differences between the cases of CPL and LPL1
is the relation between $\sigma_{yx}^{\textrm{C}}$ and $\sigma_{xy}^{\textrm{C}}$.
Figure \ref{fig2}(d) shows the $u$ dependences of $\sigma_{xy}^{\textrm{C}}$
in graphene driven by CPL, LPL1, LPL2, and LPL3.
Comparing this figure with Fig. \ref{fig2}(c),
we see that
$\sigma_{xy}^{\textrm{C}}=-\sigma_{yx}^{\textrm{C}}$ for CPL,
whereas $\sigma_{xy}^{\textrm{C}}=\sigma_{yx}^{\textrm{C}}$ for LPL1.
They are the Onsager reciprocal relations~\cite{Onsager1,Onsager2},
and their difference comes from the difference in the time-reversal symmetry.
We should note that
$\sigma_{xy}^{\textrm{C}}=\sigma_{yx}^{\textrm{C}}$ for LPL1 does not contradict
the properties of the $C_{3}$ rotational symmetry (see Appendix B).
Since the anomalous Hall conductivity is off-diagonal and antisymmetric, 
our results indicate that
the light-induced AHE
comes from 
a combination of mirror symmetry breaking and time-reversal symmetry breaking. 
This is consistent with the AHE in nondriven systems~\cite{Kon-AHE1,Kon-AHE2}. 
Our results also suggest that
the off-diagonal symmetric charge conductivity can be regarded as
an indicator for mirror symmetry breaking
in the presence of the time-reversal symmetry.
This might
be used to detect helical higher-order topological insulators~\cite{Helical-HO-TI},
which are protected by the mirror symmetry and time-reversal symmetry,
because that conductivity vanishes with the mirror symmetry
or its combining symmetry, as shown above.

Another difference is about the quantization of $\sigma_{yx}^{\textrm{C}}$.
$\sigma_{yx}^{\textrm{C}}$ is quantized only with CPL.
This quantization can be understood
using a high-frequency expansion~\cite{Highw1,Highw2},
as shown in previous studies~\cite{Light-AHE1,Mikami}: 
the term proportional to $\Omega^{-1}$ gives a pure-imaginary hopping integral
between next-nearest neighbors on the honeycomb lattice,
which is similar to the term vital for the quantum Hall effect~\cite{Haldane}. 
Similarly, we can understand the non-quantized $\sigma_{yx}^{\textrm{C}}$ with LPL:
the $\Omega^{-1}$ term becomes zero.
This is consistent with the property that LPL does not break
the time-reversal symmetry.
Note that except the above interpretations,
we do not use the high-frequency expansion.

\begin{figure}
  \begin{center}
    \includegraphics[width=60mm]{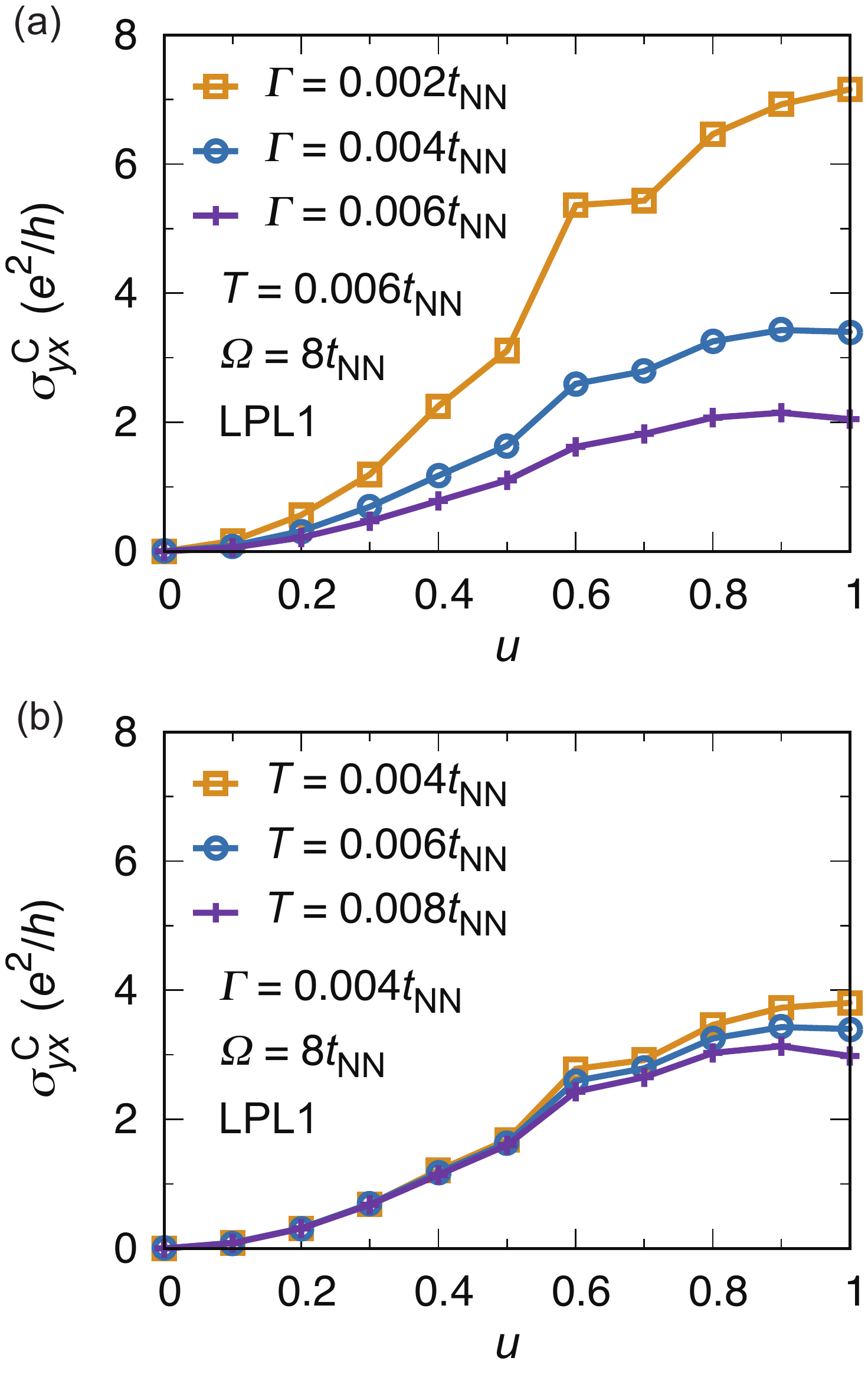}
    \end{center}
  \caption{\label{fig3}
    (Color online)
    (a), (b) The dependences of $\sigma_{yx}^{\textrm{C}}$
    on the damping induced by the system-bath coupling, $\Gamma$,
    and the temperature of the bath, $T$, 
    for graphene driven by LPL1.
  }
\end{figure}

The other difference is about the $\Gamma$ dependence of
$\sigma_{yx}^{\textrm{C}}$.
Figure \ref{fig3}(a) shows
the $\Gamma$ dependence of $\sigma_{yx}^{\textrm{C}}$
for graphene driven by LPL1.
$\sigma_{yx}^{\textrm{C}}$ is roughly proportional to $\Gamma^{-1}$.
This contrasts the $\Gamma$ dependence
of the off-diagonal charge conductivity
for graphene driven by CPL
because it is almost independent of $\Gamma$
[e.g., compare the red curve in Fig. \ref{fig2}(d) of this paper
  and the brown one in Fig. 10 of Ref.~\cite{Mikami}].
Note that $\sigma_{yx}^{\textrm{C}}$ with LPL1, as well as that with CPL,
is little dependent on the bath temperature $T$ [Fig. \ref{fig3}(b)].
This is because the bath temperature may play a similar role to
the temperature appearing in the distribution function.
We should note that $\Gamma$ is independent of temperature in our theory. 

The sign of the off-diagonal symmetric charge conductivity can be reversed 
by replacing LPL1 by the LPL for $\alpha_{x}=-\alpha_{y}=1$
or $-\alpha_{x}=\alpha_{y}=1$,
a counterpart connected by the mirror operation about
the $xz$ or $yz$ plane, respectively.
Furthermore, it remains unchanged
by replacing LPL1 by the LPL for $\alpha_{x}=\alpha_{y}=-1$.
The similar properties hold for arbitrary $\theta$
when $\alpha_{x}=\cos\theta$ and $\alpha_{y}=\sin\theta$.
These three additional results are shown in Appendix F.
They also suggest the vital role of
the mirror symmetries and their combining symmetries in
the off-diagonal charge conductivity.

\begin{figure}
  \begin{center}
    \includegraphics[width=60mm]{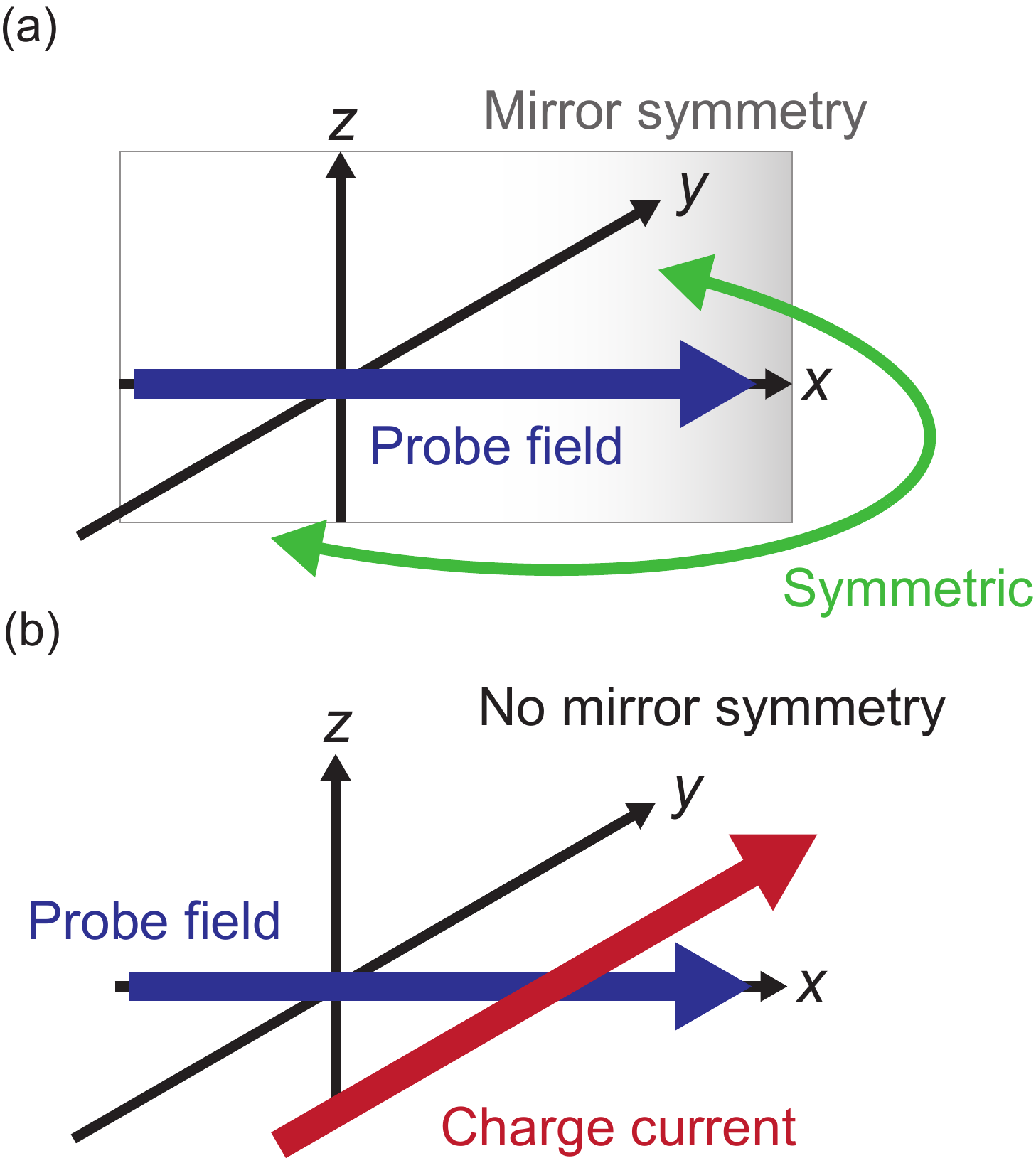}
    \end{center}
  \caption{\label{fig4}
    (Color online)
    (a), (b) Situations with and without the mirror symmetry
    (or its combining symmetry with a uniform time translation)
    under the probe field applied along the $x$ axis.
  }
\end{figure}

\section{Discussion}

The importance of the mirror symmetry breaking is a general concept.
Let us consider a situation
where the probe field 
is applied along the $x$ axis of a material.
If the mirror symmetry about the $xz$ plane
(or its combining symmetry with a uniform time translation) exists,
any currents along the $y$ axis are prohibited [Fig. \ref{fig4}(a)].
Meanwhile,
if it is broken,
the charge current along the $y$ axis can be induced [Fig. \ref{fig4}(b)].
This current is finite (i.e., $\sigma_{yx}^{\textrm{C}}\neq 0$)
if the mirror symmetry about the $yz$ plane
(or its combining symmetry),
as well as that about the $xz$ plane,
is broken.
Therefore,
the mirror symmetry breaking by light plays the key role
in the light-induced off-diagonal charge transport.
Although mirror symmetry breaking about the $xy$ plane
is important in several systems with the Rashba spin-orbit coupling~\cite{Rashba-AHE1,Rashba-AHE2},
it is not essential for obtaining
$\sigma_{yx}^{\textrm{C}}$ and $\sigma_{xy}^{\textrm{C}}$; 
such off-site spin-orbit coupling is absent in our system.
Note that the importance of the mirror symmetry breaking
can been seen from Eq. (\ref{eq:sig^C})
and the expression using the Berry curvature 
because both contain the momentum summation of the product
of the $x$ and $y$ components of the group velocity,
which can be finite without the mirror symmetries
about the $xz$ and $yz$ planes
and their combining symmetries. 

Our results can be tested experimentally. 
In our system,
the nonequilibrium steady state can be achieved
due to $\Gamma$ at times larger than
$\tau=\frac{\hbar}{2\Gamma}=O(10\textrm{fs})$.
Then,
the off-diagonal charge conductivity in graphene driven by LPL1 
could be observed experimentally in pump-probe measurements. 
Note that $u=\frac{eE_{0}a_{\textrm{NN}}}{\Omega}=0.1$
at $\Omega=8t_{\textrm{NN}}\approx 24$ eV corresponds to
$E_{0}\approx 171$ MV cm$^{-1}$.
In this estimate, 
we have used
$a_{\textrm{NN}}\approx 0.14$ nm~\cite{Graphene-LatticeConstant}. 
We have also set $t_{\textrm{NN}}\approx 3$ eV
because, according to the first-principles calculations without $\bdA(t)$~\cite{Graphene-DFT},
the energy difference between the two bands in our model at $\bdk=\bdzero$
corresponds to about $19$ eV, i.e., $6t_{\textrm{NN}}\approx 19$ eV.
Because of $t_{\textrm{NN}}\approx 3$ eV,
$k_{\textrm{B}}T=0.006t_{\textrm{NN}}(\approx 0.018$ eV$)$
corresponds to about $209$K,
where $k_{\textrm{B}}\approx 8.6\times 10^{-5}$ eV K$^{-1}$ is used.
Since $\sigma_{yx}^{\textrm{C}}$ and $\sigma_{xy}^{\textrm{C}}$ become finite at nonzero $u$'s
[Figs. \ref{fig2}(c) and \ref{fig2}(d)],
the off-diagonal charge transport induced by LPL1 is testable. 

\section{Conclusion}

We have studied the polarization dependence of
the light-induced mirror symmetry breaking and its effects on charge transport
in periodically driven graphene.
We showed that
the mirror symmetries about the $xz$ and $yz$ planes
and their combining symmetries
are broken by CPL and
by the LPL whose $A_{x}$ and $A_{y}$ are both nonzero. 
This mirror symmetry breaking leads to the light-induced AHE
in the absence of the time-reversal symmetry.
This indicates that
the origin of the light-induced AHE is
a combination of the time-reversal symmetry breaking and the mirror symmetry breaking. 
In the presence of time-reversal symmetry,
the mirror symmetry breaking results in 
the off-diagonal symmetric charge conductivity.
This conductivity could be used to
detect the mirror symmetry breaking with the time-reversal symmetry.
Our results highlight the overlooked role of the light-induced mirror symmetry breaking
in the light-induced AHE
and reveal the emergence of the off-diagonal symmetric charge transport
induced by LPL.

\begin{acknowledgments}
This work was supported by
JST CREST Grant No. JPMJCR1901, 
JSPS KAKENHI Grant No. JP22K03532, 
and MEXT Q-LEAP Grant No. JP-MXS0118067426.
\end{acknowledgments}

\appendix
\section{Light-induced mirror symmetry breaking for the Floquet Hamiltonian}
  
We analyze the polarization dependence of the light-induced mirror symmetry breaking
for the Floquet Hamiltonian.
As we will show below,
this polarization dependence is the same as that for the time-dependent Hamiltonian,
which has been shown in the main text.
(Note that the Floquet Hamiltonian is distinct from
that obtained in a high-frequency expansion.)
The momentum dependence of 
the Floquet Hamiltonian is characterized by
\begin{align}
  [\epsilon_{AB}(\bdk)]_{mn}
  &=\int_{0}^{T_{\textrm{p}}}\frac{dt}{T_{\textrm{p}}}
  e^{i(m-n)\Omega t}\epsilon_{AB}(\bdk,t)\notag\\
  &=[t_{AB}^{Z}]_{mn}e^{-ik_{y}}
  +[t_{AB}^{X}]_{mn}e^{i\frac{\sqrt{3}}{2}k_{x}}e^{i\frac{k_{y}}{2}}\notag\\
  &+[t_{AB}^{Y}]_{mn}e^{-i\frac{\sqrt{3}}{2}k_{x}}e^{i\frac{k_{y}}{2}},\label{eq:eAB_mn}\\
  [\epsilon_{BA}(\bdk)]_{mn}
  &=\int_{0}^{T_{\textrm{p}}}\frac{dt}{T_{\textrm{p}}}
  e^{i(m-n)\Omega t}\epsilon_{BA}(\bdk,t)\notag\\
  &= [t_{BA}^{Z}]_{mn}e^{ik_{y}}
  +[t_{BA}^{X}]_{mn}e^{-i\frac{\sqrt{3}}{2}k_{x}}e^{-i\frac{k_{y}}{2}}\notag\\
  &+[t_{BA}^{Y}]_{mn}e^{i\frac{\sqrt{3}}{2}k_{x}}e^{-i\frac{k_{y}}{2}},\label{eq:eBA_mn}
\end{align}
where
\begin{align}
  &[t_{ab}^{Z}]_{mn}=\int_{0}^{T_{\textrm{p}}}\frac{dt}{T_{\textrm{p}}}
  e^{i(m-n)\Omega t}t_{ab}^{Z}(t),\\
  &[t_{ab}^{X}]_{mn}=\int_{0}^{T_{\textrm{p}}}\frac{dt}{T_{\textrm{p}}}
  e^{i(m-n)\Omega t}t_{ab}^{X}(t),\\
  &[t_{ab}^{Y}]_{mn}=\int_{0}^{T_{\textrm{p}}}\frac{dt}{T_{\textrm{p}}}
  e^{i(m-n)\Omega t}t_{ab}^{Y}(t).
\end{align}
Note that 
$\epsilon_{AB}(\bdk,t)$ and $\epsilon_{BA}(\bdk,t)$ have been given by
Eqs. (\ref{eq:e_AB-App}) and (\ref{eq:e_BA-App}), respectively.
(We neglect the energy shifts due to the light frequency in the Floquet Hamiltonian
because such momentum-independent terms do not affect mirror symmetries.)
Note that we have considered
$[\epsilon_{AB}(\bdk)]_{mn}$ and $[\epsilon_{BA}(\bdk)]_{mn}$
because they give the finite components of $H_{m,n}$,
which is part of the matrix used to obtain the quasienergy in the Floquet state~\cite{Mikami}.
As we have explained in the main text,
the mirror symmetry about the $xz$ plane is preserved
if the hopping integrals as a function of time satisfy Eq. (\ref{eq:sym-xzMirror}).
Therefore,
for the Floquet Hamiltonian,
this mirror symmetry is preserved if
\begin{align}
  [t_{AB}^{Z}]_{mn}=[t_{BA}^{Z}]_{mn},\
  [t_{AB}^{Y}]_{mn}=[t_{BA}^{X}]_{mn},\
  [t_{AB}^{X}]_{mn}=[t_{BA}^{Y}]_{mn}.
\end{align}
In the case of the system driven by LPL described by Eq. (\ref{eq:A_pump-LPL}),
the hopping integrals are given by
\begin{align}
  &[t_{AB}^{Z}]_{mn}=t_{\textrm{NN}}i^{n-m}\mathcal{J}_{m-n}(\alpha_{y}u),\\
  &[t_{AB}^{X}]_{mn}=t_{\textrm{NN}}i^{n-m}
  \mathcal{J}_{n-m}(\frac{\sqrt{3}\alpha_{x}+\alpha_{y}}{2}u),\\
  &[t_{AB}^{Y}]_{mn}=t_{\textrm{NN}}i^{n-m}
  \mathcal{J}_{m-n}(\frac{\sqrt{3}\alpha_{x}-\alpha_{y}}{2}u),\\
  &[t_{BA}^{Z}]_{mn}=t_{\textrm{NN}}i^{n-m}\mathcal{J}_{n-m}(\alpha_{y}u),\\
  &[t_{BA}^{X}]_{mn}=t_{\textrm{NN}}i^{n-m}
  \mathcal{J}_{m-n}(\frac{\sqrt{3}\alpha_{x}+\alpha_{y}}{2}u),\\
  &[t_{BA}^{Y}]_{mn}=t_{\textrm{NN}}i^{n-m}
  \mathcal{J}_{n-m}(\frac{\sqrt{3}\alpha_{x}-\alpha_{y}}{2}u),
\end{align}
where $\mathcal{J}_{l}(x)$ is the Bessel function of the first kind with the order $l$. 
Therefore,
the mirror symmetry about the $xz$ plane is preserved
for the LPL with $\alpha_{x}\neq 0$, $\alpha_{y}=0$,
whereas it is broken by the LPL
with $\alpha_{x}=0$, $\alpha_{y}\neq 0$
or $\alpha_{x}\neq 0$, $\alpha_{y}\neq 0$.
Similarly, we can show that
this mirror symmetry is broken by CPL described by Eq. (\ref{eq:A_pump-CPL}). 
These results are the same as the polarization dependence shown in the main text. 

\section{$C_{3}$ rotational symmetry and mirror symmetries without or
  with a uniform time translation}

  We discuss the $C_{3}$ rotational symmetry, the mirror symmetries,
  and the combining symmetries with a uniform time translation. 
  First, we show that
  the $C_{3}$ rotational symmetry is broken with CPL or LPL,
  that the combining symmetry,
  the symmetry of a combination of the $C_{3}$ rotation and
  a uniform time translation,
  is preserved with CPL if $T_{t}: t \rightarrow t+\frac{2\pi}{3\Omega}$,
  and that the combining symmetry is broken with LPL. 
  Next, we review the mirror symmetry about the $xz$ or $yz$ plane
  with CPL or LPL,
  which has been discussed in Sect. III.
  Then,
  we show that, as well as the mirror symmetries,
  the combining symmetries, 
  the symmetries of
  combinations of the mirror operations about the $xz$ and $yz$ planes
  and a uniform time translation, 
  are broken with CPL or LPL1. 
  We also show that
  the symmetry of a combination of the mirror operation about the $yz$ or $xz$ plane
  and the uniform time translation $T_{t}: t \rightarrow t-\frac{\pi}{\Omega}$
  is preserved with LPL2 or LPL3, respectively.
  As shown in Sect. III, 
  the mirror symmetry about the $yz$ or $xz$ plane is broken
  with LPL2 or LPL3, respectively.
  Note that for LPL1 $\alpha_{x}\neq 0$ and $\alpha_{y}\neq 0$ in Eq. (\ref{eq:A_pump-LPL}),
  for LPL2 $\alpha_{x}\neq 0$ and $\alpha_{y}= 0$,
  and for LPL3 $\alpha_{x}= 0$ and $\alpha_{y}\neq 0$.
  
  We begin with the properties about the $C_{3}$ rotational symmetry
  for graphene driven by CPL.
  This periodically driven system has 
  the $C_{3}$ rotational symmetry if
  the $H_{\textrm{s}}(t)$ remains unchanged
  after a counterclockwise rotation of $120$ degrees around the $z$ axis.
  This condition can be expressed as the following equation: 
  \begin{align}
    C_{3}^{-1}H_{\textrm{s}}(t)C_{3}=H_{\textrm{s}}(t).\label{eq:C3-nonsteady}
  \end{align}
  In the case of graphene,
  this equation can be rewritten as follows:
  \begin{align}
    &C_{3}^{-1}t_{AB}^{Z}(t)C_{3}=t_{AB}^{Z}(t),\label{eq:C3-cond1}\\
    &C_{3}^{-1}t_{AB}^{X}(t)C_{3}=t_{AB}^{X}(t),\label{eq:C3-cond2}\\
    &C_{3}^{-1}t_{AB}^{Y}(t)C_{3}=t_{AB}^{Y}(t),\label{eq:C3-cond3}
  \end{align}
  where $t_{AB}^{Z}(t)=t_{\textrm{NN}}e^{-ie\bdA(t)\cdot\bdR_{0}}$,
  $t_{AB}^{X}(t)=t_{\textrm{NN}}e^{-ie\bdA(t)\cdot\bdR_{1}}$,
  and $t_{AB}^{Y}(t)=t_{\textrm{NN}}e^{-ie\bdA(t)\cdot\bdR_{2}}$.
  Since the $C_{3}$ rotation transforms
  the $Z$, $X$, and $Y$ bonds on the honeycomb lattice
  into the $X$, $Y$, and $Z$ bonds, respectively [see Fig. \ref{fig1}(a) or \ref{fig1}(b)],
  the left-hand sides of Eqs. (\ref{eq:C3-cond1}){--}(\ref{eq:C3-cond3})
  become
  \begin{align}
    &C_{3}^{-1}t_{AB}^{Z}(t)C_{3}=t_{AB}^{X}(t),\label{eq:C3-cond1_left}\\
    &C_{3}^{-1}t_{AB}^{X}(t)C_{3}=t_{AB}^{Y}(t),\label{eq:C3-cond2_left}\\
    &C_{3}^{-1}t_{AB}^{Y}(t)C_{3}=t_{AB}^{Z}(t).\label{eq:C3-cond3_left}
  \end{align}
  Therefore, if the hopping integrals satisfy
  \begin{align}
    &C_{3}^{-1}t_{AB}^{Z}(t)C_{3}=t_{AB}^{X}(t)=t_{AB}^{Z}(t),\label{eq:C3-cond1_rewrite}\\
    &C_{3}^{-1}t_{AB}^{X}(t)C_{3}=t_{AB}^{Y}(t)=t_{AB}^{X}(t),\label{eq:C3-cond2_rewrite}\\
    &C_{3}^{-1}t_{AB}^{Y}(t)C_{3}=t_{AB}^{Z}(t)=t_{AB}^{Y}(t),\label{eq:C3-cond3_rewrite} 
  \end{align}
  the $C_{3}$ rotational symmetry is preserved; otherwise, it is broken.
  Combining Eqs. (\ref{eq:C3-cond1_rewrite}){--}(\ref{eq:C3-cond3_rewrite})
  with
  the equations of the hopping integrals for graphene driven by CPL,
  \begin{align}
    &t_{AB}^{Z}(t)=t_{\textrm{NN}}e^{-iu\sin(\Omega t)},\label{eq:hopZ-CPL}\\
    &t_{AB}^{X}(t)=t_{\textrm{NN}}e^{-iu\sin(\Omega t-\frac{2\pi}{3})},\label{eq:hopX-CPL}\\
    &t_{AB}^{Y}(t)=t_{\textrm{NN}}e^{-iu\sin(\Omega t-\frac{4\pi}{3})},\label{eq:hopY-CPL}
  \end{align}
  we find that
  these hopping integrals do not satisfy
  Eqs. (\ref{eq:C3-cond1_rewrite}){--}(\ref{eq:C3-cond3_rewrite}),
  which means that 
  the $C_{3}$ rotational symmetry is broken 
  for graphene driven by CPL.
  Meanwhile,
  the periodically driven system has the combining symmetry,
  the symmetry of a combination of the $C_{3}$ rotation and
  a uniform time translation $T_{t}$, if
  \begin{align}
    T_{t}^{-1}C_{3}^{-1}H_{\textrm{s}}(t)C_{3}T_{t}=H_{\textrm{s}}(t),\label{eq:C3-steady}
  \end{align}
  which is reduced in the case of graphene to
  \begin{align}
    &T_{t}^{-1}C_{3}^{-1}t_{AB}^{Z}(t)C_{3}T_{t}
    =t_{AB}^{Z}(t),\label{eq:C3-steady-cond1}\\
    &T_{t}^{-1}C_{3}^{-1}t_{AB}^{X}(t)C_{3}T_{t}
    =t_{AB}^{X}(t),\label{eq:C3-steady-cond2}\\
    &T_{t}^{-1}C_{3}^{-1}t_{AB}^{Y}(t)C_{3}T_{t}
    =t_{AB}^{Y}(t).\label{eq:C3-steady-cond3}
  \end{align}
  If we consider the uniform time translation,
  \begin{align}
    T_{t}: t \rightarrow t+\frac{2\pi}{3\Omega},\label{eq:time-trans}
  \end{align}
  the left-hand sides of Eqs. (\ref{eq:C3-steady-cond1}){--}(\ref{eq:C3-steady-cond3})
  are written in graphene driven by CPL as follows:
  \begin{align}
    &T_{t}^{-1}C_{3}^{-1}t_{AB}^{Z}(t)C_{3}T_{t}
    =T_{t}^{-1}t_{AB}^{X}(t)T_{t}=t_{AB}^{Z}(t),\\
    &T_{t}^{-1}C_{3}^{-1}t_{AB}^{X}(t)C_{3}T_{t}
    =T_{t}^{-1}t_{AB}^{Y}(t)T_{t}=t_{AB}^{X}(t),\\
    &T_{t}^{-1}C_{3}^{-1}t_{AB}^{Y}(t)C_{3}T_{t}
    =T_{t}^{-1}t_{AB}^{Z}(t)T_{t}=t_{AB}^{Y}(t),
  \end{align}
  where we have used Eqs. (\ref{eq:C3-cond1_left}){--}(\ref{eq:C3-cond3_left})
  and (\ref{eq:hopZ-CPL}){--}(\ref{eq:hopY-CPL}).
  Therefore,
  the combining symmetry, which may be called a time-screw symmetry,
  is preserved for graphene driven by CPL.
  Because of this property,
  the time-averaged off-diagonal charge conductivities
  in the nonequilibrium steady state with CPL satisfy
  $\sigma_{xy}^{\textrm{C}}=-\sigma_{yx}^{\textrm{C}}$~\cite{Mikami}
  even without using the Onsager reciprocal relation.
  
  In a similar way, we can show for graphene driven by LPL
  that 
  the $C_{3}$ rotational symmetry and its combining symmetry are both broken.
  This result holds for arbitrary $\alpha_{x}$ and $\alpha_{y}$
  of Eq. (\ref{eq:A_pump-LPL}).
  Because of the breaking of these symmetries,
  there is no need to satisfy $\sigma_{xy}^{\textrm{C}}=-\sigma_{yx}^{\textrm{C}}$
  in the case of LPL.
  Namely, $\sigma_{xy}^{\textrm{C}}=\sigma_{yx}^{\textrm{C}}$ with LPL1 is symmetrically reasonable. 
  Note that for graphene driven by LPL
  the hopping integrals are given by
  \begin{align}
    &t_{AB}^{Z}(t)=t_{\textrm{NN}}e^{-iu\alpha_{y}\cos(\Omega t)},\label{eq:hopZ-LPL}\\
    &t_{AB}^{X}(t)=t_{\textrm{NN}}e^{iu\frac{\sqrt{3}}{2}\alpha_{x}\cos(\Omega t)}
    e^{iu\frac{\alpha_{y}}{2}\cos(\Omega t)},\label{eq:hopX-LPL}\\
    &t_{AB}^{Y}(t)=t_{\textrm{NN}}e^{-iu\frac{\sqrt{3}}{2}\alpha_{x}\cos(\Omega t)}
    e^{iu\frac{\alpha_{y}}{2}\cos(\Omega t)}.
    \label{eq:hopY-LPL}
  \end{align}

  We turn to the properties of the mirror symmetry about the $xz$ or $yz$ plane.
  The periodically driven system is symmetric with respect to
  the $xz$ mirror plane [Fig. \ref{fig1}(a)] if
  \begin{align}
    \sigma_{\textrm{m}}^{-1}H_{\textrm{s}}(t)\sigma_{\textrm{m}}=H_{\textrm{s}}(t),\label{eq:mirror1-nonsteady}
  \end{align}
  which is written in the case of graphene as
  \begin{align}
    &\sigma_{\textrm{m}}^{-1}t_{AB}^{Z}(t)\sigma_{\textrm{m}}=t_{AB}^{Z}(t),\label{eq:mirror1-cond1}\\
    &\sigma_{\textrm{m}}^{-1}t_{AB}^{X}(t)\sigma_{\textrm{m}}=t_{AB}^{X}(t),\label{eq:mirror1-cond2}\\
    &\sigma_{\textrm{m}}^{-1}t_{AB}^{Y}(t)\sigma_{\textrm{m}}=t_{AB}^{Y}(t).\label{eq:mirror1-cond3}
  \end{align}
  As we can see from Fig. \ref{fig1}(a), 
  the left-hand sides of these equations become
  \begin{align}
    &\sigma_{\textrm{m}}^{-1}t_{AB}^{Z}(t)\sigma_{\textrm{m}}=t_{BA}^{Z}(t),\label{eq:mirror1-cond1_left}\\
    &\sigma_{\textrm{m}}^{-1}t_{AB}^{X}(t)\sigma_{\textrm{m}}=t_{BA}^{Y}(t),\label{eq:mirror1-cond2_left}\\
    &\sigma_{\textrm{m}}^{-1}t_{AB}^{Y}(t)\sigma_{\textrm{m}}=t_{BA}^{X}(t).\label{eq:mirror1-cond3_left}
  \end{align}
  A combination of Eqs. (\ref{eq:mirror1-cond1}){--}(\ref{eq:mirror1-cond3})
  and Eqs. (\ref{eq:mirror1-cond1_left}){--}(\ref{eq:mirror1-cond3_left})
  gives Eq. (\ref{eq:sym-xzMirror}).
  Therefore,
  as we have shown in Sect. III,
  the mirror symmetry about the $xz$ plane is broken
  for graphene driven
  by CPL and LPL with $\alpha_{y}\neq 0$,
  whereas it is preserved for graphene driven
  by LPL with $\alpha_{y}=0$ (see Table \ref{tab}).
  For the mirror symmetry about the $yz$ plane,
  Eqs. (\ref{eq:mirror1-nonsteady}) and (\ref{eq:mirror1-cond1}){--}(\ref{eq:mirror1-cond3})
  are replaced by
  \begin{align}
    (\sigma_{\textrm{m}}^{\prime})^{-1}H_{\textrm{s}}(t)\sigma_{\textrm{m}}^{\prime}
    =H_{\textrm{s}}(t),\label{eq:mirror2-nonsteady}
  \end{align}
  and
  \begin{align}
    &(\sigma_{\textrm{m}}^{\prime})^{-1}t_{AB}^{Z}(t)\sigma_{\textrm{m}}^{\prime}
    =t_{AB}^{Z}(t),\label{eq:mirror2-cond1}\\
    &(\sigma_{\textrm{m}}^{\prime})^{-1}t_{AB}^{X}(t)\sigma_{\textrm{m}}^{\prime}
    =t_{AB}^{X}(t),\label{eq:mirror2-cond2}\\
    &(\sigma_{\textrm{m}}^{\prime})^{-1}t_{AB}^{Y}(t)\sigma_{\textrm{m}}^{\prime}
    =t_{AB}^{Y}(t).\label{eq:mirror2-cond3}
  \end{align}
  Moreover,
  as we can see from Fig. \ref{fig1}(b),
  the left-hand sides of Eqs. (\ref{eq:mirror2-cond1}){--}(\ref{eq:mirror2-cond3}) are expressed as
  \begin{align}
    &(\sigma_{\textrm{m}}^{\prime})^{-1}t_{AB}^{Z}(t)\sigma_{\textrm{m}}^{\prime}
    =t_{AB}^{Z}(t),\label{eq:mirror2-cond1_left}\\
    &(\sigma_{\textrm{m}}^{\prime})^{-1}t_{AB}^{X}(t)\sigma_{\textrm{m}}^{\prime}
    =t_{AB}^{Y}(t),\label{eq:mirror2-cond2_left}\\
    &(\sigma_{\textrm{m}}^{\prime})^{-1}t_{AB}^{Y}(t)\sigma_{\textrm{m}}^{\prime}
    =t_{AB}^{X}(t).\label{eq:mirror2-cond3_left}
  \end{align}
  Therefore,
  as we have remarked in Sect. III, 
  the mirror symmetry about the $yz$ plane is broken
  for graphene driven by CPL or LPL with $\alpha_{x}\neq 0$,
  whereas it is preserved for graphene driven
  by LPL with $\alpha_{x}=0$ (see Table \ref{tab}).
  Then,
  for the combining symmetry for the $xz$ mirror plane,
  Eqs. (\ref{eq:mirror1-nonsteady}) and (\ref{eq:mirror2-nonsteady}) are replaced by
  \begin{align}
    T_{t}^{-1}\sigma_{\textrm{m}}^{-1}H_{\textrm{s}}(t)\sigma_{\textrm{m}}T_{t}
    =H_{\textrm{s}}(t),\label{eq:mirror1-steady}
  \end{align}
  and
  \begin{align}
    T_{t}^{-1}(\sigma_{\textrm{m}}^{\prime})^{-1}H_{\textrm{s}}(t)\sigma_{\textrm{m}}^{\prime}T_{t}
    =H_{\textrm{s}}(t),\label{eq:mirror2-steady}
  \end{align}
  respectively.
  In the case of graphene,
  the former is decomposed into
  \begin{align}
    &T_{t}^{-1}\sigma_{\textrm{m}}^{-1}t_{AB}^{Z}(t)\sigma_{\textrm{m}}T_{t}
    =T_{t}^{-1}t_{BA}^{Z}(t)T_{t}
    =t_{AB}^{Z}(t),\label{eq:mirror1-cond1-steady}\\
    &T_{t}^{-1}\sigma_{\textrm{m}}^{-1}t_{AB}^{X}(t)\sigma_{\textrm{m}}T_{t}
    =T_{t}^{-1}t_{BA}^{Y}(t)T_{t}
    =t_{AB}^{X}(t),\label{eq:mirror1-cond2-steady}\\
    &T_{t}^{-1}\sigma_{\textrm{m}}^{-1}t_{AB}^{Y}(t)\sigma_{\textrm{m}}T_{t}
    =T_{t}^{-1}t_{BA}^{X}(t)T_{t}
    =t_{AB}^{Y}(t),\label{eq:mirror1-cond3-steady}
  \end{align}
  and the latter is decomposed into
  \begin{align}
    &T_{t}^{-1}(\sigma_{\textrm{m}}^{\prime})^{-1}t_{AB}^{Z}(t)\sigma_{\textrm{m}}^{\prime}T_{t}
    =T_{t}^{-1}t_{AB}^{Z}(t)T_{t}
    =t_{AB}^{Z}(t),\label{eq:mirror2-cond1-steady}\\
    &T_{t}^{-1}(\sigma_{\textrm{m}}^{\prime})^{-1}t_{AB}^{X}(t)\sigma_{\textrm{m}}^{\prime}T_{t}
    =T_{t}^{-1}t_{AB}^{Y}(t)T_{t}
    =t_{AB}^{X}(t),\label{eq:mirror2-cond2-steady}\\
    &T_{t}^{-1}(\sigma_{\textrm{m}}^{\prime})^{-1}t_{AB}^{Y}(t)\sigma_{\textrm{m}}^{\prime}T_{t}
    =T_{t}^{-1}t_{AB}^{X}(t)T_{t}
    =t_{AB}^{Y}(t).\label{eq:mirror2-cond3-steady}
  \end{align}
  In deriving the first and the last three equations,
  we have used Eqs. (\ref{eq:mirror1-cond1_left}){--}(\ref{eq:mirror1-cond3_left})
  and Eqs. (\ref{eq:mirror2-cond1_left}){--}(\ref{eq:mirror2-cond3_left}), respectively.
  Using Eqs. (\ref{eq:hopZ-CPL}){--}(\ref{eq:hopY-CPL})
  or (\ref{eq:hopZ-LPL}){--}(\ref{eq:hopY-LPL}),
  we can show that the combining symmetries
  for the $xz$ and $yz$ mirror planes are broken
  for graphene driven by CPL or LPL1.
  This is because
  in the case of graphene driven by CPL or LPL1, 
  there is no uniform time translation which makes
  the system after the mirror operation the same as
  that before it.
  For example, in the case of the mirror symmetry about the $yz$ plane with CPL,  
  if we choose $T_{t}$ in
  Eqs. (\ref{eq:mirror2-cond1-steady}){--}(\ref{eq:mirror2-cond3-steady})
  as Eq. (\ref{eq:time-trans}),
  we get
  \begin{align}
    &T_{t}^{-1}(\sigma_{\textrm{m}}^{\prime})^{-1}t_{AB}^{Z}(t)\sigma_{\textrm{m}}^{\prime}T_{t}
    =T_{t}^{-1}t_{AB}^{Z}(t)T_{t}\notag\\
    &=t_{\textrm{NN}}e^{-iu\sin(\Omega t+\frac{2\pi}{3})}
    =t_{AB}^{Y}(t)
    \neq t_{AB}^{Z}(t),\label{eq:mirror2-cond1-rewrite-steady-CPL}\\
    &T_{t}^{-1}(\sigma_{\textrm{m}}^{\prime})^{-1}t_{AB}^{X}(t)\sigma_{\textrm{m}}^{\prime}T_{t}
    =T_{t}^{-1}t_{AB}^{Y}(t)T_{t}\notag\\
    &=t_{\textrm{NN}}e^{-iu\sin(\Omega t-\frac{2\pi}{3})}
    =t_{AB}^{X}(t),\label{eq:mirror2-cond2-rewrite-steady-CPL}\\
    &T_{t}^{-1}(\sigma_{\textrm{m}}^{\prime})^{-1}t_{AB}^{Y}(t)\sigma_{\textrm{m}}^{\prime}T_{t}
    =T_{t}^{-1}t_{AB}^{X}(t)T_{t}\notag\\
    &=t_{\textrm{NN}}e^{-iu\sin(\Omega t)}
    =t_{AB}^{Z}(t)
    \neq t_{AB}^{Y}(t).\label{eq:mirror2-cond3-rewrite-steady-CPL}
  \end{align}
  There is no uniform time translation that 
  the three conditions for the hopping integrals are satisfied simultaneously.
  In contrast,
  the combining symmetry for the $yz$ or $xz$ mirror plane
  is preserved in graphene driven by LPL2 or LPL3, respectively.
  As we have shown in Sect. III,
  the mirror symmetry about the $xz$ plane
  is preserved for graphene driven by LPL2,
  whereas it is broken for graphene driven by LPL3.
  Meanwhile,
  the hopping integrals for graphene driven by LPL3 satisfy
  Eqs. (\ref{eq:mirror1-cond1-steady}){--}(\ref{eq:mirror1-cond3-steady})
  if the $T_{t}$ is chosen as
  \begin{align}
    T_{t}: t \rightarrow t-\frac{\pi}{\Omega}.\label{eq:time-trans2}
  \end{align}
  By using the same $T_{t}$,
  we can show that
  the hopping integrals for graphene driven by LPL2 satisfy
  Eqs. (\ref{eq:mirror2-cond1-steady}){--}(\ref{eq:mirror2-cond3-steady}).
  Therefore,
  the symmetry of a combination of
  the mirror operation about the $yz$ or $xz$ plane
  and the uniform time translation of Eq. (\ref{eq:time-trans2}),
  which may be called a time-glide symmetry,
  is preserved for graphene driven by LPL2 or LPL3,
  respectively.

\section{Derivation of Eq. (\ref{eq:sig^C})}

We derive Eq. (\ref{eq:sig^C}).
Since this derivation has been explained, for example, in Ref.~\cite{NA-FloquetSHE},
we explain the main points below.
Treating $\bdA_{\textrm{prob}}(t)$ in the linear-response theory,
we express a charge conductivity as
\begin{align}
  \sigma_{\mu\nu}^{\textrm{C}}(t,t^{\prime})
  &=\frac{1}{i\omega}
  \frac{\delta \langle j_{\textrm{C}}^{\mu}(t)\rangle}{\delta A_{\textrm{prob}}^{\nu}(t^{\prime})},
  \label{eq:sig-start}
\end{align}
where $\langle j_{\textrm{C}}^{\mu}(t)\rangle$ is the expectation value of
the operator of the charge current density $j_{\textrm{C}}^{\mu}(t)=J_{\textrm{C}}^{\mu}(t)/V$,
$J_{\textrm{C}}^{\mu}(t)$ is the charge current operator,
\begin{align}
  J_{\textrm{C}}^{\mu}(t)
  =(-e)\sum_{\bdk}\sum_{a,b}\sum_{\sigma=\uparrow,\downarrow}
  v_{ab}^{\mu}(\bdk,t)
  c_{\bdk a\sigma}^{\dagger}(t)c_{\bdk b\sigma}(t),\label{eq:JC^y}
\end{align}
and $v_{ab}^{\mu}(\bdk,t)=\frac{\partial \epsilon_{ab}(\bdk,t)}{\partial k_{\mu}}$.
By substituting Eq. (\ref{eq:JC^y}) into Eq. (\ref{eq:sig-start})
and doing some calculations~\cite{NA-FloquetSHE}, 
we get
\begin{align}
  \sigma_{\mu\nu}^{\textrm{C}}(t,t^{\prime})
  &=\sigma_{\mu\nu}^{\textrm{C}(1)}(t,t^{\prime})+\sigma_{\mu\nu}^{\textrm{C}(2)}(t,t^{\prime}),\label{eq:sig-divided}
\end{align}
where
\begin{align}
  \sigma_{\mu\nu}^{\textrm{C}(1)}(t,t^{\prime})
  &=\frac{e}{\omega V}\sum_{\bdk}\sum_{a,b}\sum_{\sigma=\uparrow,\downarrow}
  \frac{\delta v_{ab}^{\mu}(\bdk,t)}{\delta A_{\textrm{prob}}^{\nu}(t^{\prime})}
  G_{b\sigma a\sigma}^{<}(\bdk;t,t),\label{eq:sig^1}\\
  \sigma_{\mu\nu}^{\textrm{C}(2)}(t,t^{\prime})
  &=\frac{(-e)^{2}}{\omega V}\sum_{\bdk}\sum_{a,b,c,d}\sum_{\sigma,\sigma^{\prime}=\uparrow,\downarrow}
  v_{ab}^{\mu}(\bdk,t)v_{cd}^{\nu}(\bdk,t^{\prime})\notag\\
  &\times 
  \Bigl[G_{b\sigma c\sigma^{\prime}}^{\textrm{R}}(\bdk;t,t^{\prime})
    G_{d\sigma^{\prime} a\sigma}^{<}(\bdk;t^{\prime},t)\notag\\
    &+G_{b\sigma c\sigma^{\prime}}^{<}(\bdk;t,t^{\prime})
    G_{d\sigma^{\prime} a\sigma}^{\textrm{A}}(\bdk;t^{\prime},t)
    \Bigr],\label{eq:sig^2}
\end{align}
and the lesser, retarded, and advanced Green's functions are defined
as follows:
\begin{align}
  G_{b\sigma^{\prime} a\sigma}^{<}(\bdk;t,t^{\prime})
  &=i\langle c_{\bdk a\sigma}^{\dagger}(t^{\prime})c_{\bdk b\sigma^{\prime}}(t)\rangle,\\
  G_{a\sigma b\sigma^{\prime}}^{\textrm{R}}(\bdk;t,t^{\prime})
  &=-i\theta(t-t^{\prime})
  \langle \{c_{\bdk a\sigma}(t),c_{\bdk b\sigma^{\prime}}^{\dagger}(t^{\prime})\}\rangle,\\
  G_{a\sigma b\sigma^{\prime}}^{\textrm{A}}(\bdk;t,t^{\prime})
  &=i\theta(t^{\prime}-t)
  \langle \{c_{\bdk a\sigma}(t),c_{\bdk b\sigma^{\prime}}^{\dagger}(t^{\prime})\}\rangle.
\end{align}
Since we consider charge transport in the nonequilibrium steady state,
we introduce the time-averaged charge conductivity, 
\begin{align}
  \sigma_{\mu\nu}^{\textrm{C}}
  &=\lim_{\omega\rightarrow 0}\textrm{Re}\int_{0}^{T_{\textrm{p}}}\frac{dt_{\textrm{av}}}{T_{\textrm{p}}}
  \int_{-\infty}^{\infty}dt_{\textrm{rel}}e^{i\omega t_{\textrm{rel}}}
  \sigma_{\mu\nu}^{\textrm{C}}(t,t^{\prime}),\label{eq:t-av-sig}
\end{align}
where 
$t_{\textrm{rel}}=t-t^{\prime}$ and $t_{\textrm{av}}=(t+t^{\prime})/2$.
By combining Eq. (\ref{eq:t-av-sig}) with Eqs. (\ref{eq:sig-divided}){--}(\ref{eq:sig^2})
and performing some calculations~\cite{NA-FloquetSHE},
we obtain
\begin{align}
  &\sigma_{\mu\nu}^{\textrm{C}}
  =\frac{(-e)^{2}}{V}\sum_{\bdk}\sum_{a,b,c,d}\sum_{\sigma,\sigma^{\prime}=\uparrow,\downarrow}
  \int_{-\Omega/2}^{\Omega/2}\frac{d\omega^{\prime}}{2\pi}
  \sum_{m,l,n,q=-\infty}^{\infty}\notag\\
  &\times
  \Bigl\{  
      [v_{ab}^{\mu}(\bdk)]_{ml}
      \frac{\partial [G_{b\sigma c\sigma^{\prime}}^{\textrm{R}}(\bdk,\omega^{\prime})]_{ln}}
           {\partial \omega^{\prime}}
      [v_{cd}^{\nu}(\bdk)]_{nq}     
      [G_{d\sigma^{\prime}a\sigma}^{<}(\bdk,\omega^{\prime})]_{qm}\notag\\
    &-[v_{ab}^{\mu}(\bdk)]_{ml}
      [G_{b\sigma c\sigma^{\prime}}^{<}(\bdk,\omega^{\prime})]_{ln}
      [v_{cd}^{\nu}(\bdk)]_{nq}
      \frac{\partial [G_{d\sigma^{\prime}a\sigma}^{\textrm{A}}(\bdk,\omega^{\prime})]_{qm}}
           {\partial \omega^{\prime}}
    \Bigr\},\label{eq:sig^C-Methods}
\end{align}
where the group velocity and Green's functions in the Floquet representation
are defined as
\begin{align}
  [v_{ab}^{\mu}(\bdk)]_{mn}
  &=\int_{0}^{T_{\textrm{p}}}\frac{dt}{T_{\textrm{p}}}
  e^{i(m-n)\Omega t}v_{ab}^{\mu}(\bdk,t),\label{eq:v-Floquet}\\
  [G_{a\sigma b\sigma^{\prime}}^{r}(\bdk,\omega^{\prime})]_{mn}
  &=\int_{-\infty}^{\infty}dt_{\textrm{rel}}e^{i(\omega^{\prime}+\frac{m+n}{2}\Omega)t_{\textrm{rel}}}
  \int_{0}^{T_{\textrm{p}}}\frac{dt_{\textrm{av}}}{T_{\textrm{p}}}
  \notag\\
  &\times e^{i(m-n)\Omega t_{\textrm{av}}}
  G_{a\sigma b\sigma^{\prime}}^{r}(\bdk;t,t^{\prime}).\label{eq:G-Floquet}
\end{align}
Equation (\ref{eq:sig^C-Methods}) is equivalent to Eq. (\ref{eq:sig^C}).

\section{Dyson's equation for the Green's functions}

The Green's functions appearing in Eq. (\ref{eq:sig^C}) are determined from 
Dyson's equation in a matrix form~\cite{NA-FloquetSHE},
\begin{align}
  G=G_{0}+G_{0}\Sigma G,\label{eq:Dyson}
\end{align}
where
\begin{align}
  G=\left(
  \begin{array}{@{\,}cc@{\,}}
    G^{\textrm{R}} & G^{\textrm{K}}\\[3pt]
    0 & G^{\textrm{A}}
  \end{array}
  \right),\
  G_{0}=\left(
  \begin{array}{@{\,}cc@{\,}}
    G^{\textrm{R}}_{0} & G^{\textrm{K}}_{0}\\[3pt]
    0 & G^{\textrm{A}}_{0}
  \end{array}
  \right),\
  \Sigma=\left(
  \begin{array}{@{\,}cc@{\,}}
    \Sigma^{\textrm{R}} & \Sigma^{\textrm{K}}\\[3pt]
    0 & \Sigma^{\textrm{A}}
  \end{array} 
  \right)\label{eq:Keldysh-rep}.
\end{align}
Here $G^{\textrm{R}}$, $G^{\textrm{A}}$, and $G^{\textrm{K}}$
are the retarded, advanced, and Keldysh Green's functions with $H_{\textrm{sb}}$,
$G^{\textrm{R}}_{0}$, $G^{\textrm{A}}_{0}$, and $G^{\textrm{K}}_{0}$
are those without $H_{\textrm{sb}}$,
and $\Sigma^{\textrm{R}}$, $\Sigma^{\textrm{A}}$, and $\Sigma^{\textrm{K}}$
are the retarded, advanced, and Keldysh self-energies
due to the second-order perturbation of $H_{\textrm{sb}}$; 
the matrix $G^{\textrm{R}}$ is, for instance,
given by $G^{\textrm{R}}=([G^{\textrm{R}}_{a\sigma b\sigma^{\prime}}(\bdk,\omega)]_{mn})$,
where $a$, $b=A$, $B$,
$\sigma$, $\sigma^{\prime}=\uparrow$, $\downarrow$,
and $m$, $n=-\infty,\cdots, 0, 1, \cdots, \infty$. 
The retarded, advanced, and Keldysh components
are related to the lesser component via the relation, such as
\begin{align}
  G^{<}=\frac{1}{2}(G^{\textrm{K}}-G^{\textrm{R}}+G^{\textrm{A}}).\label{eq:G^<-relation}
\end{align}
In the second-order perturbation theory,
in which $H_{\textrm{sb}}$ is treated as perturbation,
$\Sigma^{\textrm{R}}$, $\Sigma^{\textrm{A}}$, and $\Sigma^{\textrm{K}}$
are given by~\cite{NA-FloquetSHE}
\begin{align}
  [\Sigma^{\textrm{R}}_{a\sigma b\sigma^{\prime}}(\bdk,\omega)]_{mn}
  &=
  -i\delta_{m,n}\delta_{a,b}\delta_{\sigma,\sigma^{\prime}}\Gamma,\label{eq:Sig^R}\\
  [\Sigma^{\textrm{A}}_{a\sigma b\sigma^{\prime}}(\bdk,\omega)]_{mn}
  &=+i\delta_{m,n}\delta_{a,b}\delta_{\sigma,\sigma^{\prime}}\Gamma,\label{eq:Sig^A}\\
  [\Sigma^{\textrm{K}}_{a\sigma b\sigma^{\prime}}(\bdk,\omega)]_{mn}
  &=-2i\Gamma\delta_{m,n}\delta_{a,b}\delta_{\sigma,\sigma^{\prime}}
  \tanh\frac{\omega+m\Omega}{2T},\label{eq:Sig^K}
\end{align}
where $\Gamma$ is the damping.
Then,
the matrices $G^{\textrm{R}}$, $G^{\textrm{A}}$, and $G^{\textrm{K}}$
can be determined from the following relations~\cite{NA-FloquetSHE}:
\begin{align}
  &(G^{\textrm{R}})^{-1}=(G^{-1})^{\textrm{R}},\label{eq:G^R-inv}\\
  &(G^{\textrm{A}})^{-1}=(G^{-1})^{\textrm{A}},\label{eq:G^A-inv}\\
  &G^{\textrm{K}}=-G^{\textrm{R}}(G^{-1})^{\textrm{K}}G^{\textrm{A}},\label{eq:G^K-inv}
\end{align}
where
\begin{align}
  G^{-1}=\left(
  \begin{array}{@{\,}cc@{\,}}
    (G^{-1})^{\textrm{R}} & (G^{-1})^{\textrm{K}}\\[3pt]
    0 & (G^{-1})^{\textrm{A}}
  \end{array}
  \right).\label{eq:G^-1}
\end{align}
Therefore,
we obtain the retarded and advanced Green's functions with $H_{\textrm{sb}}$
using Eqs. (\ref{eq:G^R-inv}) and (\ref{eq:G^A-inv}) with the equations, 
\begin{align}
  [(G^{-1})^{\textrm{R}}_{a\sigma b\sigma^{\prime}}(\bdk,\omega)]_{mn}
  &=(\omega+m\Omega+i\Gamma)\delta_{m,n}\delta_{a,b}\delta_{\sigma,\sigma^{\prime}}\notag\\
  &-[\epsilon_{ab}(\bdk)]_{mn}\delta_{\sigma,\sigma^{\prime}},\\
  [(G^{-1})^{\textrm{A}}_{a\sigma b\sigma^{\prime}}(\bdk,\omega)]_{mn}
  &=(\omega+m\Omega-i\Gamma)\delta_{m,n}\delta_{a,b}\delta_{\sigma,\sigma^{\prime}}\notag\\
  &-[\epsilon_{ab}(\bdk)]_{mn}\delta_{\sigma,\sigma^{\prime}},
\end{align}
where
\begin{align}
  [\epsilon_{ab}(\bdk)]_{mn}
  =\int_{0}^{T_{\textrm{p}}}\frac{dt}{T_{\textrm{p}}}e^{i(m-n)\Omega t}
  \epsilon_{ab}(\bdk,t).\label{eq:energy-dis}
\end{align}
We also get 
the Keldysh Green's function with $H_{\textrm{sb}}$
using these Green's functions, Eq. (\ref{eq:G^K-inv}), and 
\begin{align}
  [(G^{-1})^{\textrm{K}}_{a\sigma b\sigma^{\prime}}(\bdk,\omega)&]_{mn}
  =2i\Gamma\delta_{m,n}\delta_{a,b}\delta_{\sigma,\sigma^{\prime}}
  \tanh\frac{\omega+m\Omega}{2T}.
\end{align}
Then, 
using these three Green's functions and Eq. (\ref{eq:G^<-relation}),
we obtain the lesser Green's function with $H_{\textrm{sb}}$.

\section{Details of the numerical calculations}

We numerically calculate $\sigma_{\mu\nu}^{\textrm{C}}$ of Eq. (\ref{eq:sig^C})
in the following procedure. 
To calculate the momentum summation,
we set $\bdk=\frac{m_{1}}{N_{1}}\bdb_{1}+\frac{m_{2}}{N_{2}}\bdb_{2}$
and $N_{1}=N_{2}=360$,
where $0\leq m_{1}<N_{1}$, $0\leq m_{2}<N_{2}$,
$\bdb_{1}={}^{t}(\frac{2\pi}{\sqrt{3}}\ \frac{2\pi}{3})$,
$\bdb_{2}={}^{t}(\frac{2\pi}{\sqrt{3}}\ -\frac{2\pi}{3})$,
and $N_{1}N_{2}=\frac{N}{2}$. 
We calculated the frequency integral
using $\int_{-\Omega/2}^{\Omega/2}d\omega^{\prime}F(\omega^{\prime})\approx \sum_{s=0}^{W-1}\Delta\omega^{\prime} F(\omega^{\prime}_{s})$,
where $\omega^{\prime}_{s}=-\Omega/2+s\Delta\omega^{\prime}$,
$\omega^{\prime}_{W}=\Omega/2$,
and 
$\Delta\omega^{\prime}=0.001t_{\textrm{NN}}$. 
Then,
to calculate the frequency derivatives of the Green's functions, 
we used
$\frac{\partial F(\omega^{\prime})}{\partial \omega^{\prime}}
\approx \frac{F(\omega^{\prime}+\Delta\omega^{\prime})-F(\omega^{\prime}-\Delta\omega^{\prime})}{2\Delta\omega^{\prime}}$.
We took the trace over the Floquet states
[i.e., $\textrm{tr}(ABCD)=\sum_{m,l,n,q=-\infty}^{\infty}A_{ml}B_{ln}C_{nq}D_{qm}$], 
replaced the summation over the Floquet indices, $\sum_{m,l,n,q=-\infty}^{\infty}$,
by $\sum_{m,l,n,q=-n_{\textrm{max}}}^{n_{\textrm{max}}}$,
and set $n_{\textrm{max}}=2$.

\section{Additional numerical results}

\begin{figure*}
  \includegraphics[width=160mm]{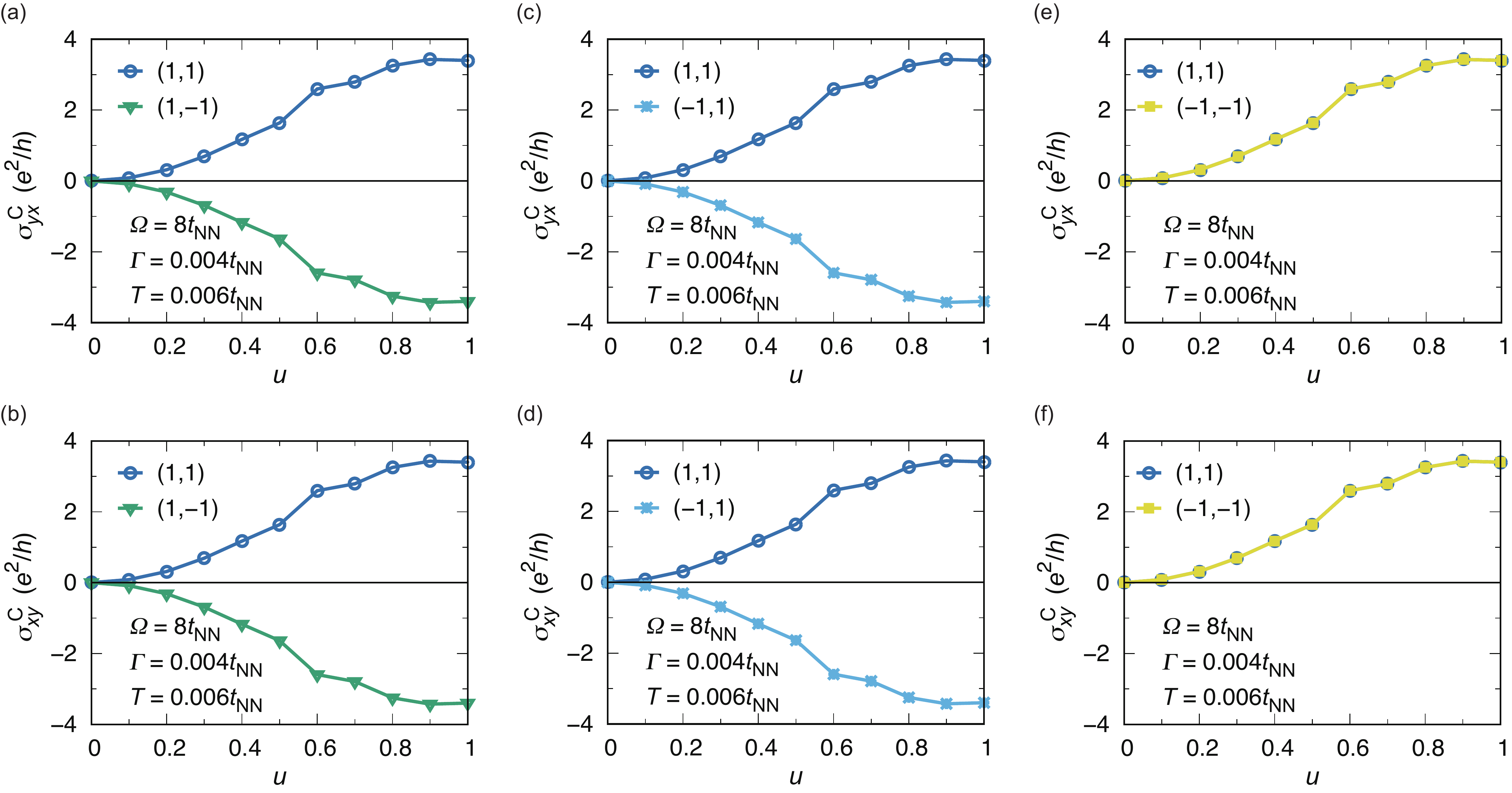}
  \caption{\label{fig5}
  (Color online)  
  (a){--}(f) The polarization dependences of
  $\sigma_{yx}^{\textrm{C}}$ and $\sigma_{xy}^{\textrm{C}}$ as functions of $u=eA_{0}$
  for graphene driven by LPL.
  The blue, green, light blue, and yellow lines correspond
  to the cases with the LPL
  for $\alpha_{x}=\alpha_{y}=1$,
  for $\alpha_{x}=-\alpha_{y}=1$,
  for $-\alpha_{x}=\alpha_{y}=1$,
  and for $\alpha_{x}=\alpha_{y}=-1$, respectively.
  }
\end{figure*}

\begin{figure*}
  \includegraphics[width=150mm]{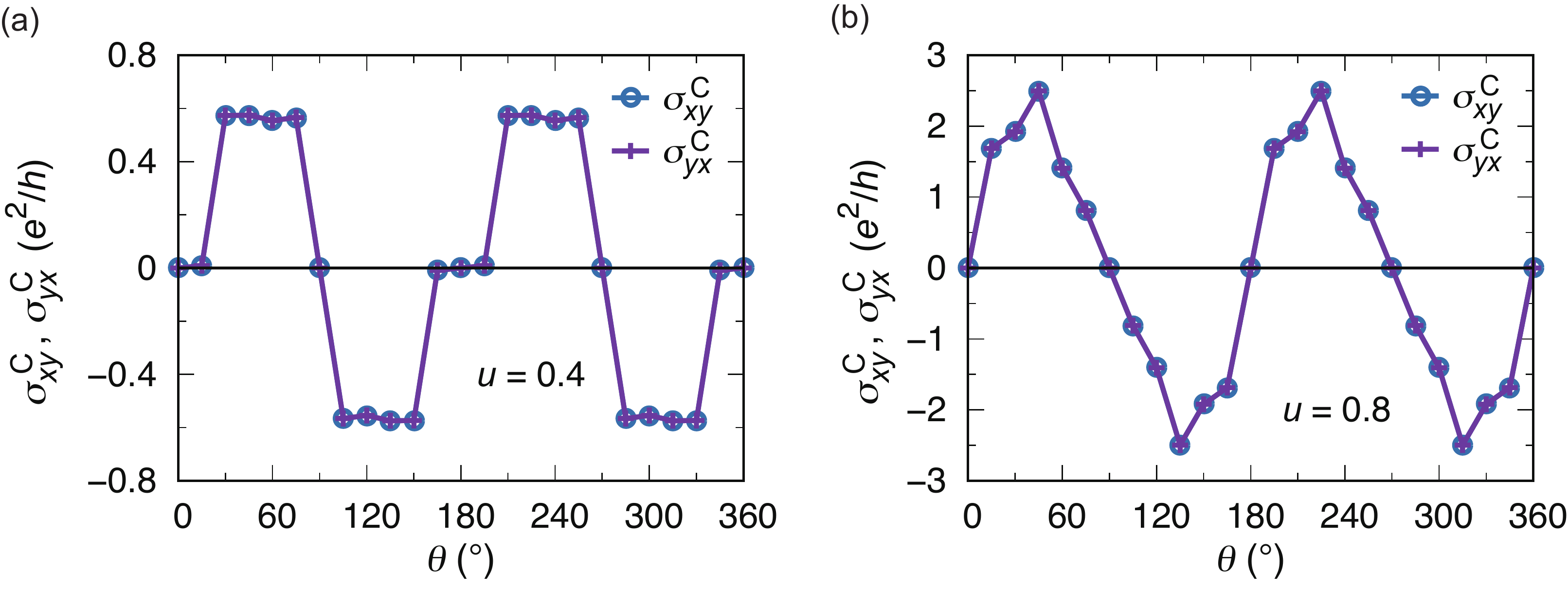}
  \caption{\label{fig6}
  (Color online)
  The $\theta$ dependences of $\sigma_{yx}^{\textrm{C}}$ and $\sigma_{xy}^{\textrm{C}}$
  at (a) $u=0.4$ and (b) $u=0.8$
  in graphene driven by LPL
  for $\alpha_{x}=\cos\theta$ and $\alpha_{y}=\sin\theta$.
  In these panels, 
  $\Omega=8t_{\textrm{NN}}$, $\Gamma=0.004t_{\textrm{NN}}$, and $T=0.006t_{\textrm{NN}}$.
  }
\end{figure*}

We show additional numerical results.
Figure \ref{fig5}(a) compares the $u$ dependences of $\sigma_{yx}^{\textrm{C}}$
in graphene driven by the LPL for $\alpha_{x}=\alpha_{y}=1$,
which has been considered as the case of LPL1 in the main text, 
and by the LPL for $\alpha_{x}=-\alpha_{y}=1$.
(Note that in both cases the mirror symmetries 
about the $xz$ and $yz$ planes
and their combining symmetries are both broken,
which means that the latter LPL also belongs to LPL1.)
The $\sigma_{yx}^{\textrm{C}}$'s in these two cases
are the same in magnitude and opposite in sign.
In addition,
$\sigma_{yx}^{\textrm{C}}=\sigma_{xy}^{\textrm{C}}$ holds
in both cases, as shown in Fig. \ref{fig5}(b).
We should note that
the systems driven by the LPL for $\alpha_{x}=\alpha_{y}=1$
and by the LPL for $\alpha_{x}=-\alpha_{y}=1$
are connected by a mirror operation with respect to the $xz$ plane,
which interchanges parts of the system above and below the $xz$ mirror plane
of Fig. \ref{fig1}(a). 
This is because
this mirror operation replaces $t_{AB}^{Z}(t)$, $t_{AB}^{Y}(t)$, and $t_{AB}^{X}(t)$
by $t_{BA}^{Z}(t)$, $t_{BA}^{X}(t)$, and $t_{BA}^{Y}(t)$, respectively
[see Eq. (\ref{eq:sym-xzMirror})]
and $t_{BA}^{Z}(t)$, $t_{BA}^{X}(t)$, and $t_{BA}^{Y}(t)$ in the former case are the same
as $t_{AB}^{Z}(t)$, $t_{AB}^{Y}(t)$, and $t_{AB}^{X}(t)$ in the latter case. 
Note that
these hopping integrals with the LPL for $\alpha_{x}=\alpha_{y}=1$ are given by
\begin{align}
  t_{AB}^{Z}(t)
  &=t_{\textrm{NN}}e^{-iu\cos\Omega t},\notag\\ 
  t_{AB}^{X}(t)
  &=t_{\textrm{NN}}e^{iu\frac{\sqrt{3}}{2}\cos\Omega t}e^{iu\frac{1}{2}\cos\Omega t},\notag\\
  t_{AB}^{Y}(t)
  &=t_{\textrm{NN}}e^{-iu\frac{\sqrt{3}}{2}\cos\Omega t}e^{iu\frac{1}{2}\cos\Omega t},\\
  t_{BA}^{Z}(t)
  &=t_{\textrm{NN}}e^{iu\cos\Omega t},\notag\\
  t_{BA}^{X}(t)
  &=t_{\textrm{NN}}e^{-iu\frac{\sqrt{3}}{2}\cos\Omega t}e^{-iu\frac{1}{2}\cos\Omega t},\notag\\
  t_{BA}^{Y}(t)
  &=t_{\textrm{NN}}e^{iu\frac{\sqrt{3}}{2}\cos\Omega t}e^{-iu\frac{1}{2}\cos\Omega t},
\end{align}
whereas those with the LPL for $\alpha_{x}=-\alpha_{y}=1$ are given by
\begin{align}
  t_{AB}^{Z}(t)
  &=t_{\textrm{NN}}e^{iu\cos\Omega t},\notag\\ 
  t_{AB}^{X}(t)
  &=t_{\textrm{NN}}e^{iu\frac{\sqrt{3}}{2}\cos\Omega t}e^{-iu\frac{1}{2}\cos\Omega t},\notag\\
  t_{AB}^{Y}(t)
  &=t_{\textrm{NN}}e^{-iu\frac{\sqrt{3}}{2}\cos\Omega t}e^{-iu\frac{1}{2}\cos\Omega t},\\
  t_{BA}^{Z}(t)
  &=t_{\textrm{NN}}e^{-iu\cos\Omega t},\notag\\
  t_{BA}^{X}(t)
  &=t_{\textrm{NN}}e^{-iu\frac{\sqrt{3}}{2}\cos\Omega t}e^{iu\frac{1}{2}\cos\Omega t},\notag\\
  t_{BA}^{Y}(t)
  &=t_{\textrm{NN}}e^{iu\frac{\sqrt{3}}{2}\cos\Omega t}e^{iu\frac{1}{2}\cos\Omega t}.
\end{align}
Moreover,
Figs. \ref{fig5}(c) and \ref{fig5}(d) show that
the sign of $\sigma_{yx}^{\textrm{C}}(=\sigma_{xy}^{\textrm{C}})$ can be reversed
by changing from the LPL for $\alpha_{x}=\alpha_{y}=1$
to that for $-\alpha_{x}=\alpha_{y}=1$.
This result can be similarly understood
because the system driven by the LPL for $-\alpha_{x}=\alpha_{y}=1$
is a counterpart connected by the mirror operation about the $yz$ plane.
Then,
$\sigma_{yx}^{\textrm{C}}(=\sigma_{xy}^{\textrm{C}})$ driven by
the LPL for $\alpha_{x}=\alpha_{y}=-1$ becomes the same as that for $\alpha_{x}=\alpha_{y}=1$,
as shown in Figs. \ref{fig5}(e) and \ref{fig5}(f).
This is because 
the system driven by the LPL for $\alpha_{x}=\alpha_{y}=-1$
is connected to that driven by the LPL for $\alpha_{x}=-\alpha_{y}=1$
by the mirror operation about the $yz$ plane
(or to that driven by the LPL for $-\alpha_{x}=\alpha_{y}=1$
by the mirror operation about the $xz$ plane).
The similar properties hold in more general cases in which
$\alpha_{x}$ and $\alpha_{y}$ are written as
$\alpha_{x}=\cos\theta$ and $\alpha_{y}=\sin\theta$,
as shown in Figs. \ref{fig6}(a) and \ref{fig6}(b).
Namely, 
$\sigma_{yx}^{\textrm{C}}(=\sigma_{xy}^{\textrm{C}})$'s driven by the LPL
for $\theta=\theta_{0}$ and $\theta_{0}+180^{\circ}$
(e.g., $\theta=30^{\circ}$ and $210^{\circ}$) are the same in magnitude and sign,
whereas those for $\theta=360^{\circ}-\theta_{0}$ and $180^{\circ}-\theta_{0}$
(e.g., $\theta=330^{\circ}$ and $150^{\circ}$)
have the opposite sign to $\theta=\theta_{0}$ and $\theta_{0}+180^{\circ}$
and the same magnitude;
these properties can be understood in a similar way.
Therefore,
these results indicate that
the sign of the off-diagonal symmetric charge conductivity can be changed 
by switching LPL1 to a counterpart connected by the mirror operation
about the $xz$ or $yz$ plane.
Note that
the relation between the LPL for $\alpha_{x}=\alpha_{y}=1$ and for $\alpha_{x}=-\alpha_{y}=1$
is similar to that between left- and right-handed circularly polarized light
because the systems driven by left- and right-handed circularly polarized light
are connected by the mirror operation
(and also by a time-reversal operation~\cite{NA-FloquetSHE}).

\end{document}